\documentclass[a4,12pt]{article}%
\usepackage{graphicx}
\usepackage{amsmath}
\usepackage{amsfonts}
\usepackage{amssymb}
\usepackage{epic}%
\newtheorem{theorem}{Theorem}[section]

\newtheorem{corollary}[theorem]{Corollary}

\newtheorem{definition}[theorem]{Definition}

\newtheorem{lemma}[theorem]{Lemma}

\newtheorem{proposition}[theorem]{Proposition}

\newenvironment{proof}[1][Proof]{\textbf{#1.} }{\ \rule{0.5em}{0.5em}}
\renewcommand{\theequation}{\thesection.\arabic{equation}}

\usepackage{epsfig}
\begin{document}

\title{Surface transitions of the semi-infinite Potts model II: the low
bulk temperature regime}

\author{C. Dobrovolny\( ^{1} \), L. Laanait\( ^{2} \), and J. Ruiz\( ^{3} \) }
\date { }
\maketitle
\renewcommand{\theequation}{\thesection.\arabic{equation}}

\renewcommand{\thefootnote}{} %
\footnote{Preprint CPT--2003/P.4570, published in J.  Stat. Phys.
\textbf{111}, 1405--1434 (2004)} 
\renewcommand{\thefootnote}{\arabic{footnote}}

\footnotetext[1]{CPT, CNRS, Luminy case 907, F-13288 Marseille Cedex
9, France.

E-mail: \textit{dodrovol@cpt.univ-mrs.fr}}

\footnotetext[2]{Ecole Normale sup\'{e}rieure de Rabat, B.P. 5118
Rabat, Morocco \\
 E-mail: \textit{laanait@yahoo.fr}}

\footnotetext[3]{CPT, CNRS, Luminy case 907, F-13288 Marseille Cedex
9, France.\\
 E-mail: \textit{ruiz@cpt.univ-mrs.fr}}

\setcounter{footnote}{3} \thispagestyle{empty}
\begin{quote}
{\footnotesize \textsc{Abstract:}  We consider the semi-infinite
\( q \)--state Potts model. We prove, for large \( q \), the existence
of a first order surface phase transition between the ordered phase
and the the so-called ``new low temperature phase'' predicted in \cite{Li},
in which the bulk is ordered whereas the surface is disordered.}

\vskip15pt

\end{quote}
\newpage

\section{Introduction and definitions}

\setcounter{equation}{0}

\subsection{Introduction}

This paper is the continuation of our study of surface phase transitions
of the semi--infinite Potts model \cite{DLR} (to be referred as paper
I).

Semi--infinite models exhibit a great variety of critical phenomena
and we refer the reader to Ref.\ \cite{Binder} for a review on this subject.

We consider, the \( q \)--states Potts model on the half-infinite
lattice with bulk coupling constant \( J \) and surface coupling
constant \( K \) (see (\ref{eq:1.1}) below for the definition of
the Hamiltonian).

Besides its popularity, this model  presents very interesting features.
Namely, in the many component limit \( q\to \infty  \), the mean field theory
yields by looking at the behavior of a bulk and a surface order parameter,
and after a suitable rescaling i.e.\ by taking the inverse temperature
\( \beta =\ln q \), the phase diagram shown in Figure~1  \cite{Li}.

\begin{center}

\setlength{\unitlength}{6.5mm} \begin{picture}(14,7) \put(3,-1){
\begin{picture}(0,0)
\drawline(0,0)(0,7)
\drawline(0,0)(8,0)       
\put(7.8,-0.15){\(\blacktriangleright\)} \put(8,-0.7){\(J\)}
\put(-0.175,6.9){\(\blacktriangle\)} \put(-1,7){\(K\)}
\put(1.9,-0.7){\(\frac{1}{d}\)} \put(5.9,-0.7){\(1\)}
\put(-1.1,2.85){\(\frac{1}{d-1}\)}
\drawline(0,3)(2,3) \drawline(2,2)(6,0) \drawline(2,0)(2,6.5)
\put(.5,1.5){I} \put(.5,4.5){II} \put(4.2,3.7){IV}
\put(2.5,0.5){III}

%\put(2.3,2.1){\(S_{2}\)} \put(2.3,3.1){\(S_{1}\)}
\end{picture}
}
\end{picture}

\end{center}

\vspace{1cm}

\begin{center}

\footnotesize{FIGURE 1: Mean field diagram borrowed from Ref.~\cite{Li}.}

\end{center}

In region (I) (respectively (IV)) the bulk spins and the surface spins
are disordered (respectively ordered). In region (II) the surface
spins are ordered while the bulk spins are disordered. The region
(III) called \textit{new low temperature phase} \cite{Li} 
corresponds to disordered
surface spins and ordered bulk spins: this phase which is also predicted
by renormalization group scheme, actually does not appear in the Ising
case \cite{FP}. On the separating line between (I) and (IV) an ordinary
transition occurs whereas so-called extraordinary phase transitions
take place on the separating lines (I)-(III) and (II)-(IV). Finally,
on the two remaining separation lines (I)-(II) and (III)-(IV), surface
phase transitions arise.

In paper I, we studied the high bulk temperature regime showing that
the first surface phase transition between a disordered and an ordered
surface while the bulk is disordered  holds whenever 
\( e^{\beta J}-1<q^{1/d} \)
and \( q \) is large enough. 

We are here concerned with the more interesting situation
in which the bulk is ordered.
We prove that
 the second surface transition between the new low temperature phase
and the ordered phase actually occurs  whenever
 \( e^{\beta J}-1>q^{1/d} \), again for
large values of \( q \).

The results are based on the analysis of the induced effect of the
bulk on the surface. 
Intuitively, this effect might be viewed as an
external magnetic field. 
When the bulk is completely ordered (a situation
that can be obtained by letting the coupling constant between bulk
sites tends to infinity) the system reduces to Potts model in dimension
\( d-1 \) with coupling constant \( K \) submitted to a magnetic
field of strength \( J \). 
Such a model is known to undergo a order-disordered
phase transition near the line \( \beta J(d-1)+\beta K=\ln q \) \cite{BBL}.
We control here this effect up to \( e^{\beta J}-1>q^{1/d} \) by
a suitable study of a surface free energy and its derivative with
respect to the surface coupling constant, which contains the thermodynamic
of the surface phase transition under consideration.

The technical tools involved in the analysis are the Fortuin-Kasteleyn
representation \cite{FK}, cluster-expansion \cite{GMM,KP,D,M}, Pirogov-Sinai
theory \cite{S}, as already in paper I, but in addition Alexander's
duality \cite{Al,L,LMeR2,PV}.

The use of Fortuin-Kasteleyn representation is two-fold. 
It provides
a uniform formulation of Ising/Potts/percolation models for which
much (but not all) of the physical theory are best implemented (see
\cite{G} for a recent review). 
It can be defined for a wide class
of model, making results easier to extend
(see e.g.\ \cite{LMR,PV,CM}). 
This representation appears in Subsection
2.1 and at the beginning of Subsection 2.2 to express both partition
functions (\( Z \) and \( Q \)) entering in the definition of the
surface free energy in terms of random cluster model.

Alexander's duality is a transformation that associates to a subcomplex
\( X \) of a cell--complex  
\( \mathbb {K} \) 
the Poincar\'e dual complex
\( [\mathbb {K}\setminus X]^{*} \) 
of its complement. 
Alexander's
Theorem provides dualities relations between the cells numbers and
Betti numbers of \( X \) and those of 
\( [\mathbb {K}\setminus X]^{*} \)
(see e.g.\ \cite{Al,L}). 
FK measures on lattices are usually expressed
in terms of the above numbers for a suitably chosen cell-complex associated
to the lattice under consideration. 
Alexander's duality provides thus
a transformation on FK configurations (and FK measures) \cite{ACCN}.
In the case of the Ising/Potts models this transformation is in fact
the counterpart of the Krammers-Wannier duality 
(or its generalizations
\cite{DW,LMeR,LMeR2}): applying it after FK gives the same result
than using first Krammers--Wannier duality and then taking FK representation
\cite{PV,BGRS}. 
We use Alexander's duality first in Subsection 2.2.
It allows to write the bulk partition function (\( Q \)) as a system
of a gas of polymers interacting through hard-core exclusion potential.
The important fact is that the activities of polymers can be controlled
for the values of parameters under consideration. 
This partition function
can then be exponentiated by standard cluster expansion. 
This duality
appears again in Subsection 2.3 to obtain a suitable expression of
the partition functions (\( Z \) ).

Cluster expansion is used again in Subsection~2.3 to express the
ratio \( Z/Q \) as a partition function of a system called Hydra
model (different from that of paper I) invariant under horizontal
translations.

Pirogov-Sinai theory, the well-known theory developed for translation
invariant systems, is then implemented in Section~3 for the study
of this system. 
Again cluster expansion enters in the game and the
needed Peierls condition is proven in Appendix.

The above description gives the organization of the paper. 
We end
this introduction with the main definitions and a statement about
the surface phase transition.

\subsection{Definitions \label{S:1.1}}

Consider a ferromagnetic Potts model on the semi-infinite lattice
\( \mathbb {L}=\mathbb {Z}^{d-1}\times \mathbb {Z}^{+} \) 
of dimension
\( d\geq 3 \). 
At each site 
\( i=\left\{ i_{1},...,i_{d}\right\} \in \mathbb {L} \),
with 
\( i_{\alpha }\in \mathbb {Z} \) 
for 
\( \alpha =1,...,d-1 \)
and 
\( i_{d}\in \mathbb {Z}^{+} \), 
there is a spin variable \( \sigma _{i} \)
taking its values in the set 
\( \mathcal{Q}\equiv \{0,1,\ldots ,q-1\} \).
We let 
\( d(i,j)=\max _{\alpha =1,...,d}\left| i_{\alpha }-j_{\alpha }\right|  \)
be the distance between two sites, 
\( d(i,\Omega )=\min _{j\in \Omega }d(i,j) \)
be the distance between the site 
\( i \) and a subset 
\( \Omega \subset \mathbb {L} \),
and 
\( d(\Omega ,\Omega ^{\prime })
=\min _{i\in \Omega ,j\in \Omega ^{\prime }}d(i,j) \)
be the distance between two subsets of 
\( \mathbb {L} \). 
The Hamiltonian
of the system is given by 
\begin{equation}
\label{eq:1.1}
H\equiv -\sum _{\langle i,j\rangle }K_{ij}\delta (\sigma _{i},\sigma _{j})
\end{equation}
 where the sum runs over nearest neighbor pairs 
\( \langle i,j\rangle  \)
(i.e.\ at Euclidean distance \( d_{\text {E}}(i,j)=1 \)) 
of a finite
subset 
\( \Omega \subset \mathbb {L} \), 
and \( \delta  \) is the
Kronecker symbol: 
\( \delta (\sigma _{i},\sigma _{j})=1 \) 
if \( \sigma _{i}=\sigma _{j} \),
and 
\( 0 \) 
otherwise. 
The coupling constants \( K_{ij} \) can take
two values according both \( i \) and \( j \) belong to the 
\emph{boundary layer} 
\( \mathbb {L}_{0}\equiv \{i\in \mathbb {L}\mid i_{d}=0\} \),
or not: 
\begin{equation}
\label{eq:1.2}
K_{ij}=\left\{ 
\begin{array}{l}
K>0\hspace {0.35cm}\text {if}\quad \langle i,j\rangle \subset \mathbb {L}_{0}
\\
J>0\hspace {0.35cm}\text {otherwise}
\end{array}\right.
\end{equation}

The partition function is defined by: 
\begin{equation}
\label{eq:1.3}
Z^{p}(\Omega )\equiv \sum e^{-\beta H}\chi _{\Omega }^{p}
\end{equation}
 Here the sum is over configurations 
\( \sigma _{\Omega }\in \mathcal{Q}^{\Omega } \),
\( \beta  \) is the inverse temperature, and \( \chi _{\Omega }^{p} \)
is a characteristic function giving the boundary conditions. 
In particular,
we will consider the following boundary conditions:

\begin{itemize}
\item 
the ordered boundary condition: 
\( \chi _{\Omega }^{\text {o}}
=\prod _{i\in \partial \Omega }\delta (\sigma _{i},0) \),
where the boundary of \( \Omega  \) 
is the set of sites of \( \Omega  \)
at distance one to its complement 
\( \partial \Omega =\left\{ i\in \Omega :d(i,\mathbb {L}\setminus \Omega )=1\right\}  \).
\item 
the ordered boundary condition in the bulk and free boundary condition
on the surface: 
\( \chi _{\Omega }^{\text {of}}
=\prod _{i\in \partial _{b}\Omega }\delta (\sigma _{i},0) \),
where
\( \partial _{b}\Omega =\partial \Omega 
\cap (\mathbb {L}\setminus \mathbb {L}_{0}) \).
\end{itemize}
Let us now consider the finite box 
\[
\Omega =
\{i\in \mathbb {L}\mid \max _{\alpha =1,...,d-1}|i_{\alpha }|\leq L,\
; 0\leq i_{d}\leq M\}
\]
 its projection 
\( \Sigma =\Omega \cap \mathbb {L}_{0}=\{i\in \Omega \mid i_{d}=0\} \)
on the boundary layer and its bulk part 
\( \Lambda =\Omega \backslash \Sigma 
=\{i\in \Omega \mid1 \leq i_{d}\leq M\} \).

The \emph{ordered surface free energy}, is defined by 
\begin{equation}
\label{eq:1.5}
g_{\text {o}}
=-\lim _{L\rightarrow \infty }\frac{1}{|\Sigma |}\lim _{M\rightarrow \infty }
\ln \frac{Z^{\text {o}}(\Omega )}{Q^{\text {o}}(\Lambda )}
\end{equation}
 Here 
\( |\Sigma |=(2L+1)^{d-1} \) 
is the number of lattice sites
in 
\( \Sigma  \), and 
\( Q^{\text {o}}(\Lambda ) \) 
is the following
bulk partition function: 
\arraycolsep2pt
\[
Q^{\text {o}}(\Lambda )
=\sum \exp \Big \{\beta J\sum _{\langle i,j\rangle \subset \Lambda }
\delta (\sigma _{i},\sigma _{j})\Big \}
\prod _{i\in \partial \Lambda }\delta (\sigma _{i},0)
\]
 where the sum is over configurations 
\( \sigma _{\Lambda }\in \mathcal{Q}^{\Lambda } \).
The surface free energy does not depend on the boundary condition
on the surface, in particular one can replace 
\( Z^{\text {o}}(\Omega ) \)
by \( Z^{\text {of}}(\Omega ) \) 
in (\ref{eq:1.5}). 
The partial
derivative of the surface free energy with respect to 
\( \beta K \)
represents the mean surface energy. 
As a result of this paper we get
for \( q \) large and \( q^{1/d}<e^{\beta J}-1<q \) 
that the mean
surface energy 
\( \frac{\partial }{\partial \beta K}g_{\text {o}} \)
is discontinuous near 
\( \beta K
=\ln \left( 1+\left( \frac{q}{e^{\beta J}-1}\right)^{1/(d-1)}\right)  \).

Namely, let 
\( \langle \, \cdot\,  \rangle ^{p} \) denote the infinite
volume expectation corresponding to the boundary condition \( p \):
\[
\langle \, f\, \rangle ^{p}(\beta J,\beta K)
=\lim _{L\rightarrow \infty ,M\rightarrow \infty }\frac{1}{Z^{p}(\Omega )}
\sum _{\sigma _{\Omega }\in \mathcal{Q}^{\Omega }}
f\, e^{-\beta H}\chi _{\Omega }^{p}
\]
 defined for local observable \( f \) and let \( e^{-\tau } \) be
defined by (\ref{eq:3.23}) below. 
As a consequence of our main result
(Theorem \ref{T:unicity} in Section 3), we have the following

\begin{corollary}
Assume that \( q^{1/d}<e^{\beta J}-1<q \) and \( q \)
is large enough, then there exists a unique value \( K_{t}(\beta ,J,q,d) \)
such that for any n.n. pair \( ij \) of the surface or between the
surface and the first layer
 \begin{alignat*}{3}
\langle\delta(\sigma_{i},\sigma_{j})\rangle^{\text{of}}(\beta J,\beta K)  &
\leq O(e^{-\tau})\quad &\text{for}\quad K\leq K_{t}
\\
\langle\delta(\sigma_{i},\sigma_{j})\rangle^{\text{o}}(\beta J,\beta K)  &
\geq1-O(e^{-\tau})\quad &\text{for}\quad K\geq K_{t}
\end{alignat*}
 \end{corollary}

In that theorem the ratios of the partition functions entering in
the definition of the surface free energy 
\( g_{\text {o}} \) (with
both \( Z^{\text {o}}(\Omega ) \) and \( Z^{\text {of}}(\Omega ) \))
are expressed in terms of partition functions of gas of polymers interacting
through a two-body hard-core exclusion potential. 
For \( q^{1/d}<e^{\beta J}-1<q \)
and \( q \) large, the associated activities are small according
the values of \( K \) namely for \( K\geq K_{t} \) with the ordered
boundary condition and for \( K\leq K_{t} \) with the ordered-free
boundary condition. 
The system is then controlled by convergent cluster
expansion.

\section{Random cluster models and Hydra model}

\setcounter{equation}{0}

\subsection{The Fortuin--Kasteleyn (FK) representation}

By using the expansion 
\( e^{\beta K_{ij}\delta (\sigma _{i},\sigma _{j})}
=1+(e^{\beta K_{ij}}-1)\delta (\sigma _{i},\sigma _{j}) \),
we obtain the Fortuin--Kasteleyn representation \cite{FK} of the
partition function: 
\begin{equation}
\label{eq:2.1}
Z^{p}(\Omega )
=\sum _{X\subset B(\Omega )}\prod _{\langle i,j\rangle \in X}
(e^{\beta K_{ij}}-1)q^{N_{\Omega }^{p}(X)}
\end{equation}
 where \( B(\Omega )=\{\langle i,j\rangle :i\in \Omega ,j\in \Omega \} \)
is the set of bonds with both endpoints belonging to \( \Omega  \),
and \( N_{\Omega }^{p}(X) \) is the number of connected components
(regarding an isolated site \( i\in \Omega  \) as a component) of
a given \( X\subset B(\Omega ) \). These numbers depend on the considered
boundary condition; introducing \( S(X) \) as the set of sites that
belong to some bond of \( X \) and \( C(X|V) \) as the number of
connected components (single sites are not included) of \( X \) that
do not intersect the set of sites \( V \), they are given by:
\begin{eqnarray*}
N_{\Omega }^{\text {o}}(X) 
& = & |\Omega |-|S(X)\cup \partial \Omega |+C(X|\partial \Omega )
\\
N_{\Omega }^{\text {of}}(X) 
& = & |\Omega |-|S(X)\cup \partial _{b}\Omega |+C(X|\partial _{b}\Omega )
\end{eqnarray*}
Hereafter \( |E| \) denotes the number of elements of the set \( E \).

We introduce the parameters
\begin{equation}
\label{eq:2.3}
\left\{ 
\begin{array}{rl}
\displaystyle  & \beta _{s}\equiv 
\displaystyle \frac{\ln (e^{\beta K}-1)}{\ln q}\\
 & \beta _{b}\equiv 
\displaystyle \frac{\ln (e^{\beta J}-1)}{\ln q}
\end{array}
\right.
\end{equation}
 and let 
\( X_{s}=X\cap B(\mathbb {L}_{0}) \), 
\( X_{b}=X\setminus X_{s} \),
to get 
\begin{equation}
\label{eq:2.4}
Z^{p}(\Omega )
=\sum _{X\subset B(\Omega )}
q^{\beta _{s}|X_{s}|+\beta _{b}|X_{b}|+N_{\Omega }^{p}(X)}
\end{equation}

The ground state diagram of this system is analogous to the diagram
of Figure~1, by replacing \( J \) by \( \beta _{b} \) and \( K \)
by \( \beta _{s} \) (see paper I).

For the bulk partition function 
\( Q^{\text {o}}(\Lambda ) \), 
one find that the FK representation reads
\begin{equation}
Q^{\text {o}}(\Lambda )
=
\sum _{Y\subset B(\Lambda )}
q^{\beta _{b}|Y|+N_{\Lambda }^{\text {o}}(Y)}
=
q^{\beta _{b}|B(\Lambda )|}
\sum _{Y\subset B(\Lambda )}
q^{-\beta _{b}|B(\Lambda )\setminus Y|+N_{\Lambda }^{\text {o}}(Y)}
\end{equation}
 where 
\( 
N_{\Lambda }^{\text {o}}(Y)=\left| \Lambda \right| 
-\left| S(X)\cup \partial \Lambda \right| +C(X|\partial \Lambda ) 
\).

\subsection{Low temperature expansion of the bulk partition function\label{expansion}}

We give in this subsection an expansion of the partition function
\( Q^{\text {o}}(\Lambda ) \) 
at 
\textquotedblleft temperature\textquotedblright\ 
\( \beta _{b}>\frac{1}{d} \).
The expansion is mainly based on a duality property and we first recall
geometrical results on Poincar\'{e} and Alexander duality 
(see e.g. \cite{L},\cite{Al},\cite{DW},\cite{KLMR}).

We first consider the lattice \( \mathbb {Z}^{d} \) and the associated
cell-complex 
\( \mathbf{L} \) 
whose objects 
\( s_{p} \) 
are called
\( p \)--cells 
(\( 0\leq p\leq d \)): 
\( 0 \)--cells are vertices,
\( 1 \)--cells are bonds, 
\( 2 \)--cells are plaquettes etc...: 
a
\( p \)--cell may be represented as 
\( (x;\sigma _{1}e_{1},...,\sigma _{p}e_{p}) \)
where 
\( x\in \mathbb {Z}^{d},(e_{1},...,e_{d}) \) 
is an orthonormal
base of 
\( \mathbb {R}^{d} \) and 
\( \sigma _{\alpha }=\pm 1,\alpha =1,...,d \).
Consider also the dual lattice 
\[
(\mathbb {Z}^{d})^{\ast }
=
\left\{ x=(x_{1}+\frac{1}{2},...,x_{d}+\frac{1}{2})
:
x_{\alpha }\in \mathbb {Z},\alpha =1,...,d\right\} 
\]
 and the associated cell complex 
\( \mathbf{L}^{\ast } \). 
There
is a one to-one correspondence 
\begin{equation}
\label{eq:2.23}
s_{p}\leftrightarrow s_{d-p}^{\ast }
\end{equation}
 between \( p \)--cells of the complex 
\( \mathbf{L} \) 
and the 
\( d-p \)--cells
of \( \mathbf{L}^{\ast } \). 
In particular to each bond 
\( s_{1} \)
corresponds the hypercube 
\( s_{d-1}^{\ast } \) that crosses 
\( s_{1} \)
in its middle. 
The dual 
\( E^{\ast } \) 
of a subset 
\( E\subset \mathbf{L} \)
is the subset of element of 
\( \mathbf{L}^{\ast } \) 
that are in
the one-to-one correspondence (\ref{eq:2.23}) with the elements of
\( E \).

We now turn to the Alexander duality in the particular case under
consideration in this paper. 
Let \( Y\subset B(\Lambda ) \) 
be a
set of bonds. 
We define the A-dual of 
\( Y \) 
as 
\begin{equation}
\widehat{Y}
=\left( B(\Lambda )\setminus Y\right)^{\ast }
\end{equation}
 As a property of Alexander duality one has
\begin{eqnarray}
\left| \widehat{Y}\right|  
& = & 
\left| B(\Lambda )\setminus Y\right| 
\\ 
N_{\Lambda }^{\text {o}}(Y)
& = & N_{\text {cl}}(\widehat{Y})
\end{eqnarray}
where \( N_{\text {cl}}(\widehat{Y}) \) 
denote the number of independent
closed (\( d-1 \))--surfaces of \( \widehat{Y} \). 
We thus get
\begin{equation}
\label{bulkpf}
Q^{\text {o}}(\Lambda )
=q^{\beta _{b}|B(\Lambda )|}
\sum_{\widehat{Y}\subset \left[ B(\Lambda )\right]^{\ast }}
q^{-\beta _{b}|\widehat{Y}|+N_{\text {cl}}(\widehat{Y})}
\end{equation}

This system can be described by a gas of polymers interacting through
hard core exclusion potential. 
Indeed, we introduce polymers as connected
subsets (in the 
\( \mathbb {R}^{d} \) 
sense) of \( (d-1) \)-cells
of 
\( \mathbf{L}^{\ast } \) 
and let 
\( \mathcal{P}(\Lambda ) \)
denote the set of polymers whose 
\( (d-1) \)--cells belong to 
\( \left[ B(\Lambda )\right] ^{\ast } \).
Two polymers 
\( \gamma _{1} \) and \( \gamma _{2} \) 
are compatible
(we will write \( \gamma _{1}\thicksim \gamma _{2} \)) 
if they do
not intersect and incompatible otherwise 
(we will write \( \gamma _{1}\nsim \gamma _{2} \)).
A family of polymers is said compatible if any two polymers of the
family are compatible and we will use 
\( \mathbf{P}(\Lambda ) \)
to denote the set of compatible families of polymers 
\( \gamma \in \mathcal{P}(\Lambda ) \).
Introducing the activity of polymers by 
\begin{equation}
\varphi _{\text {o}}(\gamma )
=q^{-\beta _{b}|\gamma |+N_{\text {cl}}(\gamma )}
\end{equation}
 one has: 
\begin{equation}
Q^{\text {o}}(\Lambda )
=q^{\beta _{b}|B(\Lambda )|}
\sum _{\widehat{Y}\in \widehat{\mathbf{P}}(\Lambda )}
\prod _{\gamma \in \widehat{Y}}\varphi _{\text {o}}(\gamma )
\end{equation}
 with the sum running over compatible families of polymers including
the empty-set with weight equal to \( 1 \).

We will now introduce multi-indexes in order to write the logarithm
of this partition function as a sum over these multi-indexes (see
\cite{M}). 
A multi-index \( C \) is a function from the set 
\( \mathcal{P}(\Lambda ) \)
into the set of non negative integers, 
and we let 
\(\text{supp} \, C=\left\{ \gamma \in \mathcal{P}(\Lambda ):C(\gamma )\geq1 \right\}  \).
We define the truncated functional 
\begin{equation}
\label{eq:2.10}
\Phi _{0}(C)=\frac{a(C)}{\prod _{\gamma }C(\gamma )!}
\prod _{\gamma }\varphi _{\text {o}}(\gamma )^{C(\gamma )}
\end{equation}
 where the factor \( a(C) \) 
is a combinatoric factor defined in
terms of the connectivity properties of the graph \( G(C) \) with
vertices corresponding to \( \gamma \in \text{supp} \, C \) (there
are \( C(\gamma ) \) vertices for each \( \gamma \in  \) 
supp\( \, C \)
) that are connected by an edge whenever the corresponding polymers
are incompatible). 
Namely, \( a(C)=0 \) and hence \( \Phi _{0}(C)=0 \)
unless \( G(C) \) is a connected graph in which case \( C \) is
called a \emph{cluster} and 
\begin{equation}
\label{eq:2.11}
a(C)=\sum _{G\subset G(C)}(-1)^{\left| e(G)\right| }
\end{equation}
 Here the sum goes over connected subgraphs \( G \) whose vertices
coincide with the vertices of \( G(C) \) and 
\( \left| e(G)\right|  \)
is the number of edges of the graph \( G \). 
If the cluster \( C \)
contains only one polymer, then \( a(\gamma )=1 \). 
In other words,
the set of all cells of polymers belonging to a cluster \( C \) is
connected. 
The support of a cluster is thus a polymer and it is then
convenient to define the following new truncated functional 
\begin{equation}
\label{eq:2.12}
\Phi (\gamma )=\sum _{C:\text{supp}\, C=\gamma }\Phi _{0}(C)
\end{equation}

As proved in paper I, we have the following
\begin{theorem}\label{T:CE}
Assume that \( \beta _{b}>1/d \) and 
\( c_{0}\nu _{d}q^{-\beta _{b}+\frac{1}{d}}\leq 1 \),
where 
\( \nu _{d}=d^{2}2^{4(d-1)} \), 
and 
\( 
c_{0}
=\left[ 1+2^{d-2}(1+\sqrt{1+2^{3-d}})\right] 
\exp \left[ \frac{2}{1+\sqrt{1+2^{3-d}}}\right]  
\),
then 
\begin{equation}
Q^{\text {o}}(\Lambda )
=
e^{\beta _{b}|B(\Lambda )|}
\exp \left\{ \sum _{\gamma \in \mathcal{P}(\Lambda )}\Phi (\gamma )
\right\}
\end{equation}
 with a sum running over (non-empty) polymers, and the truncated functional
\( \Phi  \) 
satisfies the estimates
\begin{equation}
\left| \Phi (\gamma )\right| 
\leq \left| \gamma \right| 
\left( c_{0}\nu _{d}
q^{-\beta _{b}+\frac{1}{d}}\right) ^{\left| \gamma \right| }
\end{equation}
\end{theorem}

The proof uses that the activities satisfy the bound 
\( \varphi _{\text {o}}(\gamma )\leq q^{-(\beta _{b}-1/d)|\gamma |} \)
(because \( N_{\text {cl}}(\gamma )\leq |\gamma |/d \)) 
and the standard
cluster expansion. 
The details are given in Ref.\ \cite{DLR}.

\subsection{Hydra model}
We now turn to the partition function \( Z^{p}(\Omega ) \). 
We will,
as in the previous subsection, apply Alexander duality. 
It will turn
out that the ratio 
\( Z^{p}(\Omega )/Q^{\text {o}}(\Lambda ) \) 
of
the partition functions  entering in the definition (\ref{eq:1.5})
of the surface free energy \( g_{\text {o}} \) 
can be expressed as
a partition function of geometrical objects to be called \emph{hydras}.

Namely, we define the A-dual of a set of bonds \( X\subset B(\Omega ) \)
as 
\begin{equation}
\widehat{X}
=\left( B(\Omega )\setminus X\right) ^{\ast }
\end{equation}
This transformation can be analogously defined in terms of the occupation
numbers 
\begin{equation}
\label{eq:DA1}
n_{b}=
\left\{ 
\begin{array}{cl}
1 & \mathrm{if}\quad b\in X
\\
0 & \mathrm{otherwise}
\end{array}
\right.
\end{equation}
For a configuration 
\( n=\{n_{b}\}_{b\in B(\Omega )}\subset \{0,1\}^{B(\Omega )} \)
we associate the configurations 
\( \widehat{n}=\{\widehat{n}_{s}\}_{s\in \left[ B(\Omega )\right] ^{\ast }}
\subset \{0,1\}^{\left[ B(\Omega )\right] ^{\ast }} \)
given
by 
\begin{equation}
\label{eq:DA2}
\widehat{n}_{b^{\ast }}=1-n_{b},\quad b\in B(\Omega )
\end{equation}
 where \( b^{\ast } \) is the \( (d-1) \)--cell dual of \( b \)
under the correspondence (\ref{eq:2.23}); (see Figure~2).

\vspace{.5cm}

\begin{center}
\epsfig{file=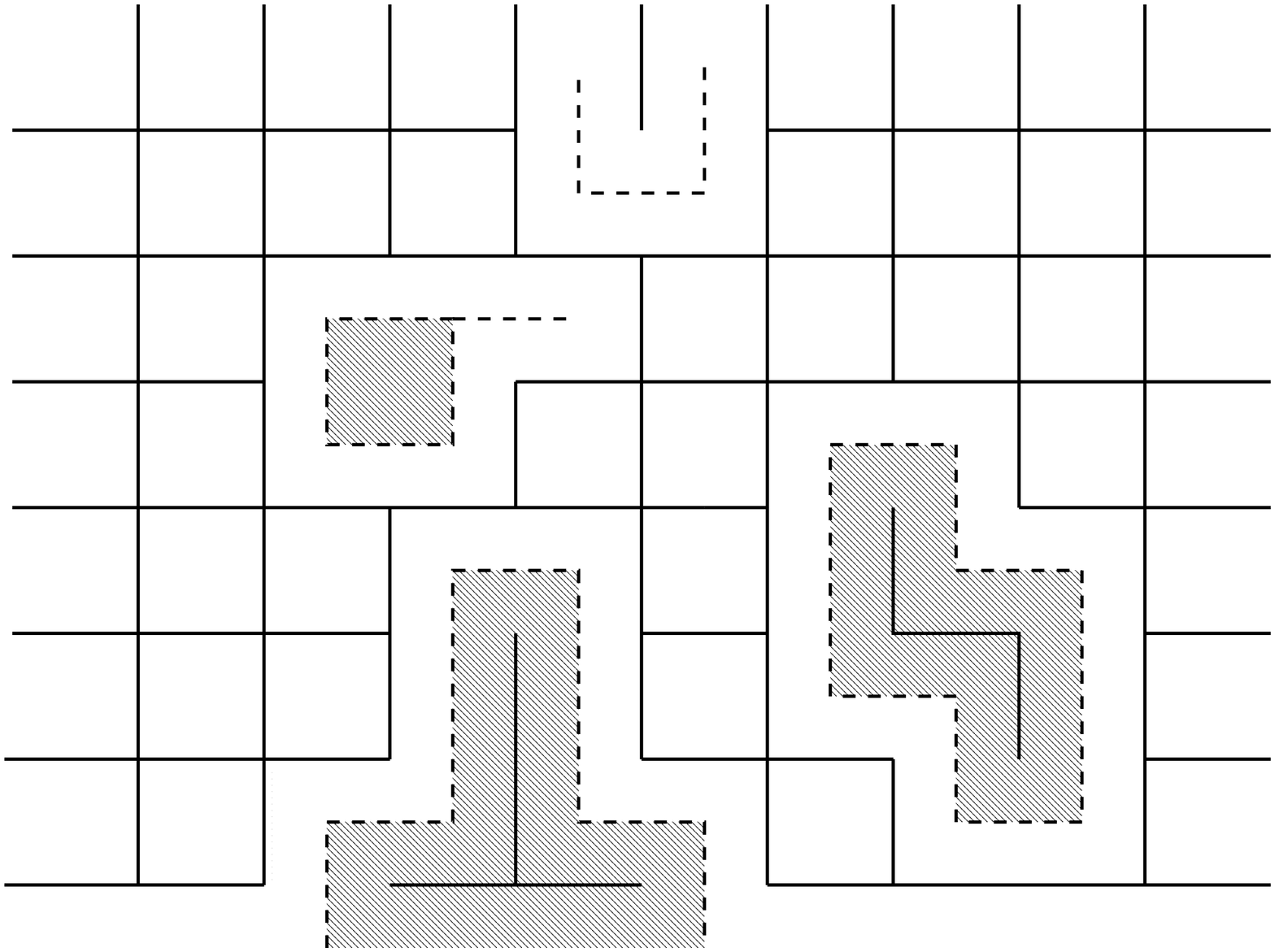,height=5cm,width=5cm}
\end{center}

\begin{center}
\footnotesize{
FIGURE 2: A configuration \( X \) (full lines) and its A-dual \( \widehat{X} \)
(dashed lines).
}
\end{center}

For any set of cells \( \widehat{X} \) 
we will use the decomposition
\( \widehat{X}=\widehat{X}_{s}\cup \widehat{Z}_{b}\cup \widehat{Y}_{b} \)
where 
\( \widehat{X}_{s} \) is the set of cells whose dual are bonds
with two endpoints on the boundary surface \( \Sigma  \), 
\( \widehat{Z}_{b} \)
is the set of cells whose dual are bonds with one endpoint on the
boundary surface 
\( \Sigma  \) 
and one endpoint in the bulk 
\( \Lambda  \)
and the remaining 
\( \widehat{Y}_{b} \) 
is the set of cells whose
dual are bonds with two endpoints in the bulk
\( \Lambda  \). 
Thus,
for the decomposition 
\( X=X_{s}\cup X_{b} \) 
introduced above, we
have 
\begin{eqnarray*}
\left| \widehat{X}_{s}\right|  
& = & \left| B(\Sigma )\setminus X_{s}\right| 
\\
\left| \widehat{Z}_{b}\right| +\left| \widehat{Y}_{b}\right|  
& = & \
\left| B(\Omega )\setminus B(\Sigma )\right| -\left| X_{b}\right|
\end{eqnarray*}
 We let \( B_{0} \) 
be the set of bonds that have an endpoint on
the boundary layer 
\( \mathbb {L}_{0} \) and the other endpoint on
the layer \( \mathbb {L}_{-1}\equiv \{i\in \mathbb {L}\mid i_{d}=-1\} \) 
and
let 
\( \widetilde{N}_{\text {cl}}(\widehat{X}) \) 
be the number of
independent closed surfaces of 
\( 
\widehat{X}\cup B_{0}^{\ast } 
\):
\( 
\widetilde{N}_{\text {cl}}(\widehat{X})
=
N_{\text {cl}}(\widehat{X}\cup B_{0}^{\ast }) 
\).
As a result of Alexander duality, 
one has
\[
N_{\Omega }^{\text {o}}(X)=\widetilde{N}_{\text {cl}}(\widehat{X})
\]
 Denoting by 
\( B_{1}(\Omega ) \) 
the set bonds that have an endpoint
in \( \partial _{s}\Omega  \) 
the other endpoint in 
\( \mathbb {L}\setminus \Omega  \),
we have furthermore
\[
N_{\Omega }^{\text {of}}(X)
=\widetilde{N}_{\text {cl}}(\widehat{X}\cup \left[ B_{1}(\Omega )\right] ^{\ast })
\]
 These formula lead to the following expression for the partition
function (\ref{eq:2.4})
\[
Z^{p}(\Omega )
=
q^{\beta _{s}|B(\Sigma )|+\beta _{b}\left| B(\Omega )\setminus B(\Sigma )\right| }
\sum _{\widehat{X}\subset \left[ B(\Omega )\right]^{\ast }}
q^{-\beta _{s}|\widehat{X}_{s}|-\beta _{b}(\left| \widehat{Z}_{b}\right| 
+\left| \widehat{Y}_{b}\right| )+\widetilde{N}_{\text {cl}}(\widehat{X})+\widehat{\chi }_{\Omega }^{p}}
\]
 where 
\( \widehat{\chi }_{\Omega }^{\text {o}}=0 \) 
and 
\( \widehat{\chi }_{\Omega }^{\text {of}}
=
\widetilde{N}_{\text {cl}}(\widehat{X}\cup \left[ B_{1}(\Omega )\right] ^{\ast })
-\widetilde{N}_{\text {cl}}(\widehat{X}) 
\).
Notice that this Boltzmann weight equals the Boltzmann weight of the
bulk partition function (\ref{bulkpf}) for those 
\( \widehat{X}\subset \left[ B(\Lambda )\right]^{\ast } \)
i.e.\ 
if 
\( \widehat{X}_{s}=\emptyset  \) 
and 
\( \widehat{Z}_{b}=\emptyset  \).
They can thus be factorized in the ratio of the two partition functions.
Namely, we define hydras as components of \( (d-1) \)--cells
not completely included in \( \left[ B(\Lambda )\right] ^{\ast } \).
\begin{definition}
A connected set of \( (d-1) \)--cells 
\( \delta \subset \left[ B(\Omega )\right] ^{\ast } \)
(in the \( \mathbb {R}^{d} \) sense) is called \emph{hydra} in \( \Omega  \),
if it contains a cell whose dual is a bond with at least one endpoint
on the boundary surface \( \Sigma  \).
\end{definition}
\begin{definition}
Given an hydra 
\( \delta \subset \left[ B(\Omega )\right] ^{\ast } \),
the components of 
\( \delta  \) 
included in 
\( \left[ B(\Sigma )\right]^{\ast } \)
are called \emph{legs} of the hydra, the components included in 
\( \left[ B(\Lambda )\right]^{\ast } \)
are called \emph{heads} of the hydra and the remaining components
are called \emph{bodies} of the hydra.
\end{definition}
The dual cells of bodies of hydras are bonds between the boundary
layer and the first layer 
\( \mathbb {L}_{1}\equiv \{i\in \mathbb {L}\mid i_{d}=1\} \);  (see Figure~3).

\vspace{-1cm}

\begin{center}

\setlength{\unitlength}{8 mm} \begin{picture}(17,6)(-3,0)
\drawline(1,0)(1,1) \drawline(2,0)(2,1)
\drawline(6,0)(6,1) \drawline(8,0)(8,1)
\drawline(10,0)(10,1)
\dashline{.1}(1,1)(3,1)
\dashline{.1}(6,1)(10,1)
\dottedline{.1}(0,1)(1,1)\dottedline{.1}(0,2)(1,2)
\dottedline{.1}(1,1)(1,2) \dottedline{.1}(0,1)(0,2)
\dottedline{.1}(1,2)(1,3)
\dottedline{.1}(2,1)(2,2) \dottedline{.1}(2,1)(2,2)
\dottedline{.1}(2,2)(2,3) \dottedline{.1}(2,3)(4,3)
\dottedline{.1}(4,2)(4,3)
\dottedline{.1}(2,2)(6,2) \dottedline{.1}(6,1)(6,2)
\dottedline{.1}(5,2)(5,3) \dottedline{.1}(5,3)(6,3)
\dottedline{.1}(6,3)(6,4)
\dottedline{.1}(9,1)(9,3)
\dottedline{.1}(9,3)(11,3)\dottedline{.1}(9,2)(10,2)
\dottedline{.1}(10,2)(10,3)
\end{picture}

\end{center}

\vspace{.2cm}

\begin{center}

\footnotesize{
FIGURE 3: {
\footnotesize A hydra, in two dimensions
(a dimension not considered in this paper), with \( 5 \) feet (components
of full lines), \( 2 \) bodies (components of dashed lines), and
\( 3 \) heads (components of dotted lines).
} 
}

\end{center}

We let \( \mathcal{H}(\Omega ) \) 
denote the set of hydras in 
\( \Omega  \).
Two hydras \( \delta _{1} \) and \( \delta _{2} \) are said compatible
(
we will write 
\( \delta _{1}\thicksim \delta _{2} \)
) 
if they do
not intersect. 
A family of hydras is said compatible if any two hydras
of the family are compatible and we let 
\( \mathbf{H}(\Omega ) \)
denote the set of compatible families of hydras 
\( \delta \in \mathcal{H}(\Omega ) \).

Clearly, a connected subset of cells included in 
\( \left[ B(\Omega )\right]^{\ast } \)
is either a hydra 
\( \delta \in \mathcal{H}(\Omega ) \) 
or a polymer
\( \gamma \in \mathcal{P}(\Lambda ) \) 
(defined in Subsection \ref{expansion}).
Therefore any subset of 
\( \left[ B(\Omega )\right] ^{\ast } \) 
is
a disjoint union of a compatible family of hydras 
\( \widehat{X}\in \mathbf{H}(\Omega ) \)
with a compatible family of polymers \( \widehat{Y}\in \mathbf{P}(\Lambda ) \).

The partition function 
\( Z^{p}(\Omega ) \) 
given by (\ref{eq:2.1})
reads thus: 
\begin{eqnarray}
Z^{p}(\Omega ) 
& = & 
q^{\beta _{s}|B(\Sigma )|+\beta _{b}\left| B(\Omega )\setminus B(\Sigma )\right| }
\label{eq:3.1} 
\\
 &  & \times 
\sum _{\widehat{X}\in \mathbf{H}(\Omega )}
q^{-\beta _{s}|\widehat{X}_{s}|-\beta _{b}(\left| \widehat{Z}_{b}\right| 
+\left| \widehat{Y}_{b}\right| )+\widetilde{N}_{\text {cl}}(\widehat{X})
+\widehat{\chi }_{\Omega }^{p}}\sum _{\widehat{Y}\in \mathbf{P}(\Lambda )
:\widehat{Y}\thicksim \widehat{X}}
\prod _{\gamma \in \widehat{Y}}\varphi _{\text {o}}(\gamma )
\nonumber
\end{eqnarray}
 where the compatibility 
\( \widehat{Y}\thicksim \widehat{X} \) 
means
no component of 
\( \widehat{Y} \) 
is connected with a component of
\( \widehat{X} \).

According to Subsection \ref{expansion}, 
the last sum in the RHS
of the above formula can be exponentiated as: 
\( \exp \left\{ \sum _{\gamma \in \mathcal{P}(\Lambda );\gamma \thicksim X}\Phi (\gamma )\right\}  \).
Hence dividing the above partition function by the partition function
\( Q^{\text {o}}(\Lambda ) \) 
we get by taking into account Theorem~\ref{T:CE}:
\begin{eqnarray}
\Xi ^{p}(\Omega ) 
& \equiv  & \frac{Z^{p}(\Omega )}{Q^{\text {f}}(\Lambda )}
=q^{\beta _{s}|B(\Sigma )|
+\beta _{b}(\left| B(\Omega )\right| -\left| B(\Sigma )\right| 
-\left| B(\Lambda )\right| )}
\label{eq:3.2} 
\\
 &  & \times \sum _{\widehat{X}\in \mathbf{H}(\Omega )}
q^{-\beta _{s}|\widehat{X}_{s}|-\beta _{b}(\left| \widehat{Z}_{b}\right| 
+\left| \widehat{Y}_{b}\right| )+\widetilde{N}_{\text {cl}}(\widehat{X})
+\widehat{\chi }_{\Omega }^{p}}
\exp \left\{ -\sum _{\gamma \in \mathcal{P}(\Lambda );\gamma \nsim \widehat{X}}
\Phi (\gamma )\right\} 
\nonumber
\end{eqnarray}
 Hereafter the incompatibility \( \gamma \nsim X \) means that no
component of \( \widehat{X} \) is connected with \( \delta  \).

\( \Xi ^{p}(\Omega ) \) 
is thus the partition function of a gas of
hydras \( \widehat{X}=\{\delta _{1},\ldots ,\delta _{n}\} \) 
interacting
through hard-core exclusion potential and through a long range interaction
potential (decaying exponentially in the distance under the hypothesis
of Theorem~\ref{T:CE}) defined on the polymers of the bulk.

If we neglect this long range potential, and if we moreover restrict
to consider only hydras without head, the system of hydras will reduce
itself to a \( (d-1) \) Potts model with two-body interaction coupling
\( K \) 
and magnetic field \( J \) 
(i.e.\ with formal Hamiltonian
\( 
H=
-
\sum _{\langle i,j\rangle \subset \mathbb {L}_{0}}K\delta (\sigma _{i},\sigma _{j})
-
\sum _{\langle i,k\rangle ,i\in \mathbb {L}_{0},k\in \mathbb {L}_{1}}J\delta (\sigma _{i},0) 
\)
).
This model undergoes a temperature driven first order phase transition,
whenever q is large enough and 
\( d\geq 3 \) \cite{BBL}. 
We will
show that it is also the case for the hydra model (\ref{eq:3.2})
implementing the fact that the heads of hydras modify only weakly
their activities and that the long range interaction potential decays
exponentially (the needed assumptions are close to those of Theorem~\ref{T:CE}).
To this end it is convenient to first rewrite this potential in terms
of a model of \emph{aggregates}. 

Let us introduce the (real-valued)
functional 
\begin{equation}
\label{eq:3.3}
\Psi (\gamma )=e^{-\Phi (\gamma )}-1
\end{equation}
 defined on polymers \( \gamma \in P(\Lambda ) \). 
An aggregate \( A \)
is a family of polymers whose support, 
\( \text{supp}\, A=\cup _{\gamma \in A}\gamma  \),
is connected. 
Two aggregates \( A_{1} \) and \( A_{2} \) are said
compatible if 
\( \text {supp}\, A_{1}\cap \text {supp}\, A_{2}=\emptyset  \).
A family of aggregates is said compatible if any two aggregates of
the family are compatible and we will use 
\( \mathbf{A}(\Lambda ) \)
to denote the set of compatible families of aggregates. 
Introducing
the statistical weight of aggregates by
\begin{equation}
\label{eq:3.4}
\omega (A)=\prod _{\gamma \in A}\Psi (\gamma )
\end{equation}
 we then get:
\begin{eqnarray}
\exp \left
\{ 
-\sum_{\substack {\gamma \in \mathcal{P}(\Lambda );\gamma \nsim X}}\Phi (\gamma )\right\}  
& = & 
\prod_{\substack {\gamma \in \mathcal{P}(\Lambda );\gamma \nsim X}}(1+\Psi (\gamma ))
\nonumber
\\
 & = & \sum _{\mathcal{A}\in \mathbf{A}(\Lambda )}
\prod_{\substack {A\in \mathcal{A};A\nsim X}}\omega (A)
\label{eq:3.5}
\end{eqnarray}
where \( A \nsim X \) means that every polymer of the aggregate \( A \)
is incompatible with \( X \). 
Since the support of aggregates is
a connected set of \( (d-1) \)--cells, i.e.\  a polymer, it is convenient
(as it was done for clusters in Subsection 2.3) to sum the statistical
weights (\ref{eq:3.4}) over aggregates with same support. 
We thus
define the weight
\begin{equation}
\label{eq:3.6}
\psi (\gamma )\equiv \sum _{A:\text {supp}\, A=\gamma }\omega (A)
\end{equation}
 with \( A\nsim X \), 
to get
\begin{eqnarray}
\Xi ^{p}\left( \Omega \right)  
& = & 
q^{\beta _{s}|B(\Sigma )|
+\beta _{b}(\left| B(\Omega )\right| 
-\left| B(\Sigma )\right| -\left| B(\Lambda )\right| )}
\label{eq:3.7} 
\\
 &  & 
\times 
\sum _{\widehat{X}\in \mathbf{H}(\Omega )}
q^{-\beta _{s}|\widehat{X}_{s}|-\beta _{b}(\left| \widehat{Z}_{b}\right| 
+\left| \widehat{Y}_{b}\right| )+\widetilde{N}_{\text {cl}}(\widehat{X})+\widehat{\chi }_{\Omega }^{p}}
\sum _{\widehat{Y}\in \mathbf{P}(\Lambda )}
\prod _{\substack {\gamma \in \widehat{Y}\gamma \nsim \widehat{X}}}\psi (\gamma )
\nonumber
\end{eqnarray}

The system is thus described by two families: a compatible family
of hydras (a subset of 
\( \left[ B(\mathbb {L})\right] ^{\ast } \))
and a compatible family of polymers 
(a subset of 
\( \left[ B(\mathbb {L\diagdown L}_{0})\right] ^{\ast } \))
each of these polymers being incompatible with the family of hydras.

We will define in the next subsection the diluted partition functions
for our system. 
This partition function differs only from the ``physical'' partition
function (\ref{eq:3.7}) by a boundary term and thus both partition
functions lead to the same free energy. The recurrence relations
of
Lemma~\ref{L:I1} below, allow to expand the diluted partition functions
in term of matching signed contours interacting through hard-core
exclusion potential.

\subsection{Diluted partition functions}

Note first that even though our model is defined on a \( d \)--dimensional
box \( \Omega  \) it has a \( (d-1) \)--dimensional structure and
the highest order of the logarithm of partition functions behaves
like \( O(\left| \Sigma \right| ) \). 
It will be convenient to consider
\( \Omega  \) 
as a set of lines and its dual \( \Omega ^{\ast } \)
as a set of columns.

We let a line \( L(x) \) be a cylinder set of sites of \( \mathbb {L} \)
whose projection on the boundary layer is the site \( x \) and whose
height is less than a given number \( M \): 
\( L(x)=\{i\in \mathbb {L}\, (i_{1},...,i_{d-1})=x\in \mathbb {L}_{0},i_{d}\leq M\} \).
We let \( \mathbb {L}_{M} \) be the set of all such lines. 
The dual
of a line is called column and is thus a set of \( d \)--cells of
the complex \( \mathbf{L}^{\ast } \). 
For 
\( \Omega \subset \mathbb {L}_{M} \),
we let 
\( \Sigma =\Omega \cap \mathbb {L}_{0} \), 
be its projection
on the boundary layer, \( \Lambda =\Omega \setminus \Sigma  \) 
and
\( \left\Vert \Omega ^{\ast }\right\Vert =\left| \Sigma \right|  \)
be the number of columns of \( \Omega ^{\ast } \) (or of lines of
\( \Omega  \)).

Consider a site \( x\in \mathbb {L} \) and its dual \( d \)--cell
\( x^{\ast } \). 
We shall use \( \mathcal{E}(x^{\ast }) \) to denote
the set of \( (d-1) \)--cells of the boundary of \( x^{\ast } \)
(there are the dual cells of the bonds whose \( x \) is an endpoint). 
For
a set of \( d \)--cells \( D \), we let 
\( \mathcal{E}(D)=\cup _{x^{\ast }\in D}\mathcal{E}(x^{\ast }) \)
be the union of the boundaries of the \( d \)--cells of \( D \).

Next, it can easily be checked that the configuration 
\( (\widehat{X}^{\text {o}}=\emptyset ,\widehat{Y}=\emptyset ) \)
and the configuration 
\( (\widehat{X}^{\text {of}}
=
\left[ B(\mathbb {L}_{0})\diagdown B(\mathbb {L\diagdown L}_{0})\right] ^{\ast },\widehat{Y}=\emptyset ) 
\) 
 are
ground states of the system.

We will use 
\( \mathbf{H}^{p}(\Omega ) \) 
to denote the set of compatible
families of hydras defined on 
\( \mathcal{E}(\Omega ^{\ast })\cap  \)
\( \left[ B(\mathbb {L})\right] ^{\ast } \) 
that coincide with \( \widehat{X}^{p} \)
on 
\( \mathcal{E}(\left[ \partial \Omega \right]^{\ast }) \), 
and
use \( \mathbf{P}^{\text {dil}}\left( \Lambda \right)  \) 
to denote
the compatible families of polymers defined on 
\( \mathcal{E}(\Omega ^{\ast })\diagdown (\mathcal{E}(\Sigma^{\ast })
\cup \mathcal{E}(\left[ \partial \Omega \right] ^{\ast }) \).

For such configurations the Boltzmann weight in (\ref{eq:3.7}) reads
\[
q^{-\beta _{s}|\widehat{X}_{s}|-\beta _{b}(\left| \widehat{Z}_{b}\right| 
+\left| \widehat{Y}_{b}\right| )+\widetilde{N}_{\text {cl}}(\widehat{X})}
\]
 since for those 
\( 
\widehat{X} \in \mathbf{H}^{\text {of}}(\Omega ) 
\)
one has 
\( 
\widetilde{N}_{\text {cl}}(\widehat{X}\cup \left[ B_{1}(\Omega )\right]^{\ast })
=
\widetilde{N}_{\text {cl}}
(\widehat{X}) 
\).

We define, for (any) volume 
\( \Omega \subset \mathbb {L} \), 
the
diluted Hamiltonian of a configuration \( \widehat{X}=\widehat{X}^{p} \)
a.e., as:
\begin{equation}
\label{eq:3.9}
H_{\Omega }^{\text {dil}}(\widehat{X})
=
\sum _{x^{\ast }\in \Omega 
^{\ast }}
e_{x^{\ast }}(\widehat{X})
-\widetilde{N}_{\text {cl}}
(\widehat{X}\cap \mathcal{E}(\Omega^{\ast }) )
\end{equation}
 where the energy per cell is defined by
\[
e_{x^{\ast }}(\widehat{X})=
\begin{array}{ccc}
\frac{\beta _{s}}{2}\left| \widehat{X}_{s}
\cap \mathcal{E}(x^{\ast })\right| +\beta _{b}\left| \widehat{Z}_{b}\cap \mathcal{E}(x^{\ast })\right|  
& \text {if} & x\in \mathbb {L}_{0}
\end{array}
\]
 for the \( d \)--cells of the surface and by
\[
e_{x^{\ast }}(\widehat{X})=
\begin{array}{ccc}
\frac{\beta _{b}}{2}\left| \widehat{Y}_{b}\cap \mathcal{E}(x^{\ast })\right|  
& \text {if} & x\in \mathbb {L}\diagdown \mathbb {L}_{0}
\end{array}\]
 for the \( d \)--cells of the bulk. 

The diluted partition function is defined by 
\begin{equation}
\Xi _{p}^{\text {dil}}\left( \Omega \right) 
=\sum _{\widehat{X}\in \mathbf{H}^{p}(\Omega )}
q^{-H_{\Omega }^{\text {dil}}(X)}\sum _{Y\in \mathbf{P}^{\text {dil}}\left( \Lambda \right) }
\prod _{\substack {\gamma \in Y\gamma \nsim X}}\psi (\gamma )
\end{equation}
Up to a boundary term \( O(\partial \Sigma ) \) 
one has \( \ln \Xi ^{p}\left( \Omega \right) 
=\left[ (d-1)\beta _{s}+\beta _{b}\right] \left\Vert \Omega \right\Vert \ln q
+\ln \Xi _{p}^{\text {dil}}\left( \Omega \right)  \),
hence 
\begin{equation}
-\lim _{\Omega \uparrow \mathbb {L}}\frac{1}{\left\Vert \Omega \right\Vert }
\ln \Xi _{p}^{\text {dil}}\left( \Omega \right) =g_{\text {o}}+
\left[ (d-1)\beta _{s}+\beta _{b}
\right] \ln q
\end{equation}
 where \( \Omega \uparrow \mathbb {L} \) 
means that we take first
the limit \( M\rightarrow \mathbb {\infty } \) and then the limit
\( \Sigma \uparrow \mathbb {L}_{0} \) in the van-Hove or Fisher sense
\cite{R}.

Notice that the diluted Hamiltonian on ground states reads on set
of columns \( \Omega ^{\ast }\subset \mathbb {L}_{M}^{\ast } \):
\begin{equation}
H_{\Omega }^{\text {dil}}(\widehat{X}^{p})=e_{p}\left\Vert \Omega ^{\ast }\right\Vert
\end{equation}
 with the specific energies
\begin{equation}
\label{eq:3.12}
\begin{array}{ll}
e_{\text {o}} & =0\\
e_{\text {of}} & =(d-1)\beta _{s}+\beta _{b}-1
\end{array}
\end{equation}

\section{Surface transition in the bulk low temperature regime}

\setcounter{equation}{0}

\subsection{Contours and Peierls estimates}

We first define the contours of our system.

Let \( \Omega \subset \mathbb {L}_{M} \), \( \Omega ^{\ast } \)
its dual set and \( (\widehat{X},\widehat{Y}) \) be a configuration
of our system in 
\( \Omega  \): 
\( \widehat{X}\in \mathbf{H}^{p}(\Omega ),\widehat{Y}
\in \mathbf{P}^{\text {dil}}(\Omega ),Y\nsim X \).

A \( d \)--cell \( x^{\ast }\in \Omega ^{\ast } \) is called 
\emph{p-correct},
if \( \widehat{X} \) coincides with the ground state \( \widehat{X}^{p} \)
on the \( (d-1) \)--cells of the boundary \( \mathcal{E}(x^{\ast }) \)
of \( x^{\ast } \) 
and the intersection 
\( \widehat{Y}\cap \mathcal{E}(x^{\ast })=\emptyset  \).
A column is called p-correct if all the \( d \)--cells of the
column are p-correct.

Columns and \( d \)--cells that are not p-correct are called \emph{incorrect}.

The set of incorrect columns of a configuration 
\( (\widehat{X},\widehat{Y}) \)
is called \emph{boundary} of the configuration 
\( (\widehat{X},\widehat{Y}) \).

A triplet 
\( \Gamma =\{\text {supp}\, \Gamma ,\widehat{X}(\Gamma ),\widehat{Y}(\Gamma )\} \),
where \( \text {supp}\, \Gamma  \) 
is a maximal connected subset
of the boundary of the configuration 
\( (\widehat{X},\widehat{Y}) \)
called support of \( \Gamma  \), 
\( \widehat{X}(\Gamma ) \) the
restriction of \( \widehat{X} \) 
to the boundary \( \mathcal{E}(\text {supp}\, \Gamma ) \)
of the support of \( \Gamma  \), and 
\( \widehat{Y}(\Gamma ) \)
the restriction of 
\( \widehat{Y} \) to \( \mathcal{E}(\text {supp}\, \Gamma ) \),
is called \emph{contour} of the configuration \( (X,Y) \). 
Hereafter
a set of \( d \)--cells is called connected if the graph that joins
all the dual sites \( i,j \) of this set with 
\( d(i,j)\leq 1 \)
is connected.

A triplet 
\( \Gamma =\{\text {supp}\, \Gamma ,\widehat{X}(\Gamma ),\widehat{Y}(\Gamma )\} \),
where \( \text {supp}\, \Gamma  \) 
is a connected set of columns
is called \emph{contour} if there exists a configuration 
\( (\widehat{X},\widehat{Y}) \)\ such
that \( \Gamma  \) is a contour of \( (\widehat{X},\widehat{Y}) \).
We will use \( \left| \Gamma \right|  \) to denote the number of
incorrect cells of 
\( \text {supp}\, \Gamma  \) and \( \left\Vert \Gamma \right\Vert  \)
to denote the number of columns of 
\( \text {supp}\, \Gamma  \).

Consider the configuration having \( \Gamma  \) as unique contour;
it will be denoted 
\( (\widehat{X}^{\Gamma },\widehat{Y}^{\Gamma }) \).
Let \( L_{p}(\Gamma ) \) be the set of p-correct columns of 
\( \mathbb {L}^{*}_{M}\setminus \text {supp}\, \Gamma  \).
Obviously, either a component of \( L_{\text {o}}(\Gamma ) \) 
is
infinite or a component of \( L_{\text {of}}(\Gamma ) \) is infinite.
In the first case \( \Gamma  \) is called contour of the ordered
class or o-contour and in the second case it is called of-contour.
When \( \Gamma  \) is a p-contour (we will let \( \Gamma ^{p} \)
denote such contours) we use \( \text {Ext}\, \Gamma  \) to denote
the unique infinite component of \( L_{p}(\Gamma ) \); 
this component
is called \emph{exterior} of the contour. 
The set of remaining components
of \( L_{p}(\Gamma ) \) is denoted \( \text {Int}_{p}\Gamma  \)
and the set \( L_{m\neq p}(\Gamma ) \) is denoted 
\( \text {Int}_{m}\Gamma  \).
The union 
\( \text {Int}\Gamma 
=\text {Int}_{\text {f}}\Gamma \cup \text {Int}_{\text {fo}}\Gamma  \)
is called \emph{interior} of the contour and 
\( V(\Gamma )=\text {supp}\, \Gamma \cup \text {Int}\Gamma  \).

Two contours \( \Gamma _{1} \) and \( \Gamma _{2} \) are said compatible
if the union of their supports is not connected. 
They are mutually
compatible external contours if \( V(\Gamma _{1})\subset \text {Ext}\Gamma _{2} \)
and \( V(\Gamma _{2})\subset \text {Ext}\Gamma _{1} \).

We will use \( G(\Gamma ^{p}) \) to denote the set of configurations
having \( \Gamma ^{p} \) as unique external contour. 
The crystal
partition function is then defined by :
\begin{equation}
\label{eq:3.15}
\Xi ^{\text {cr}}(\Gamma ^{p})
=\sum _{(\widehat{X},\widehat{Y})\in G(\Gamma ^{p})}
q^{-H_{V(\Gamma ^{p})}^{\text {dil}}(\widehat{X})}
\prod _{\gamma \in Y}\psi (\gamma )
\end{equation}

\begin{lemma}
\label{L:I1}
The following set of recurrence equations holds:
\begin{equation}
\label{eq:3.16}
\Xi _{p}^{\text {dil}}(\Omega )
=\sum _{\{\Gamma _{1}^{p},\ldots ,\Gamma _{n}^{p}\}_{\text {ext}}}
q^{-e_{p}\left\Vert \text {Ext}\right\Vert }\prod _{i=1}^{n}\Xi ^{\text {cr}}(\Gamma _{i}^{p})
\end{equation}
 Here the sum is over families 
\( \{\Gamma _{1}^{p},\ldots ,\Gamma _{n}^{p}\}_{\text {ext}} \)
of mutually compatible external contours in 
\( \Omega  \) 
(\( \text {supp}\, \Gamma _{i}^{p}\subset \Omega _{\text {int}}
=\left\{ i\in \Omega :d(i,\mathbb {L}_{M}\setminus \Omega )>1\right\}  
\)),
\( \left\Vert \text {Ext}\right\Vert 
=\left\Vert \Omega ^{*}\right\Vert -\sum\limits _{i}\left\Vert V(\Gamma _{i}^{p})\right\Vert  
\) 
where
\( \left\Vert V(\Gamma _{i}^{p})\right\Vert  \) 
is
the number of columns of \( V(\Gamma _{i}^{p}) \); 
\begin{equation}
\label{eq:3.17}
\Xi ^{\text {cr}}(\Gamma ^{p})=\varrho (\Gamma ^{p})\, 
\prod _{m\in \left\{ \text {o},\text {of}\right\} }\Xi _{m}^{\text {dil}}(\text {Int}_{m}\Gamma ^{p})\,
\end{equation}
 where: 
\begin{equation}
\label{eq:3.18}
\varrho (\Gamma ^{p})\equiv 
q^{-H_{\text {supp}\, \Gamma ^{p}}^{\text {dil}}(X^{^{_{\Gamma ^{p}}}})}
\prod _{\gamma \in Y_{\Gamma ^{p}}}\psi (\gamma )
\end{equation}
\end{lemma}

\begin{proof}
We have only to observe that for any 
\( \widehat{X}\in \mathbf{H}^{p}(\Omega ) \)
\begin{equation}
H_{\Omega }^{\text {dil}}(\widehat{X})
=\sum _{\Gamma }H_{\text {supp}\, \Gamma }^{\text {dil}}(\widehat{X}^{\Gamma })
+\sum _{p}e_{p}\left\Vert L_{p}(\widehat{X})\cap \Omega ^{\ast }\right\Vert
\end{equation}
 where the sum is over all contours of the boundary of the configuration
\( (\widehat{X},\widehat{Y}=\emptyset ) \) 
and 
\( \left\Vert L_{p}(\widehat{X})\cap \Omega ^{*}\right\Vert  \)
is the number of \( p \)--correct columns inside \( \Omega  \) 
of
this configuration.
\end{proof}

Lemma \ref{L:I1} gives the following expansion for the partition
function 
\begin{equation}
\label{eq:3.20}
\Xi _{p}^{\text {dil}}(\Omega )
=q^{-e_{p}\left\Vert \Omega \right\Vert }
\sum _{_{\{\Gamma _{1}^{p},\ldots ,\Gamma _{n}^{p}\}_{_{\text {comp}}}}}
\prod _{i=1}^{n}z(\Gamma _{i}^{p})
\end{equation}
 where the sum is now over families of compatibles contours of the
same class and 
\begin{equation}
\label{eq:3.21}
z(\Gamma _{i}^{p})=\varrho (\Gamma ^{p})
q^{e_{p}\left\Vert \Gamma ^{p}\right\Vert }
\frac{
\Xi _{m}^{\text {dil}}(\text {Int}_{m}\Gamma ^{p})
}{
\Xi _{p}^{\text {dil}}(\text {Int}_{m}\Gamma ^{p})
}
\end{equation}
 where \( \left\Vert \Gamma ^{p}\right\Vert  \) is the number is
the number of columns of \( \text {supp}\, \Gamma ^{p} \) and \( m\neq p \).

To control the behavior of our system, we need to show Peierls condition,
that means that \( \varrho (\Gamma ^{p})q^{e_{p}\left\Vert \Gamma ^{p}\right\Vert } \)
has good decaying properties with respect to the number of incorrect
cells of \( \text {supp}\, \Gamma ^{p} \). We use in fact the modified
Peierls condition introduced in Ref.~\cite{KP2} 
where 
\( \varrho (\Gamma ^{p})q^{e_{p}\left\Vert \Gamma ^{p}\right\Vert } \)
is replaced by 
\( \varrho (\Gamma ^{p})q^{\underline{e}\left\Vert \Gamma ^{p}\right\Vert } \)
with 
\( \underline{e}=\min \left( e_{\text {o}},e_{\text {of}}\right)  \).

Let 
\begin{equation}
\label{eq:3.23}
e^{-\tau }=\left( 2^{(3d-2)}q^{-\frac{1-\beta _{b}}{2(d-1)}}+3c2^{d+1}
\nu _{d}^{3}q^{\frac{1}{d}-\beta _{b}}\right) ^{\left\Vert S\right\Vert }
\frac{1}{1-6c\nu _{d}^{3}q^{\frac{1}{d}-\beta _{b}}}
\end{equation}
 where 
\( c=8e(e-1)c_{0} \) 
and 
\( \nu _{d}=d^{2}2^{4(d-1)} \).
We have the following

\begin{proposition}
\label{P:Peierls}
Let 
\( S\subset \mathbb {L}_{M}^{\ast } \) 
be a finite connected
set of columns, assume that 
\( \frac{1}{d}<\beta _{b}<1 \) 
and 
\( 6c\nu _{d}^{3}q^{-\frac{1}{d}+\beta _{b}}<1 \),
then for all 
\( \beta _{s}\in \mathbb {R} \): 
\begin{equation}
\label{eq:3.24}
\sum _{\Gamma ^{p}:\text {supp}\, \Gamma ^{p}=S}
\left| \varrho (\Gamma ^{p})\right| q^{\underline{e}\left\Vert \Gamma ^{p}\right\Vert }
\leq e^{-\tau \left\Vert S\right\Vert }
\end{equation}
 where 
\( \left\Vert S\right\Vert  \) 
is the number of columns of
\( S \).
\end{proposition}

The proof is postponed to the Appendix.

The recurrence equations of Lemma \ref{L:I1} together with the Peierls
estimates (\ref{eq:3.24}) allow to study the states invariant under
horizontal translation (HTIS) of the hydra system as in paper I. 
This
is the subject of the next subsection.

\subsection{Diagram of horizontal translation invariant states}

To state our result, we first define the functional
\begin{equation}
\label{eq:3.25}
K_{p}(S)=\sum _{\Gamma ^{p}:\text {supp}\, \Gamma ^{p}=S}z(\Gamma ^{p})
\end{equation}
 Consider the partition function 
\( \Xi _{p}^{\text {dil}}(\Omega ) \)
(\ref{eq:3.20}) 
and for a compatible family 
\( \left\{ \Gamma _{1}^{p},...,\Gamma _{n}^{p}\right\} _{\text {comp}} \)
of 
\( p \)--contours, denote by \( S_{1},...,S_{n} \) 
their respective
supports. 
By summing over all contours with the same support this
partition function can be written as the partition function of a gas
of polymers \( S \) with activity 
\( K_{p}(S)=\sum\limits _{\Gamma ^{p}:\text {supp}\, \Gamma ^{p}=S}z(\Gamma ^{p}) \)
interacting through hard-core exclusion potential:
\begin{equation}
\label{eq:3.26}
\Xi _{p}^{\text {dil}}(\Omega )
=q^{-e_{p}\left\Vert \Omega \right\Vert }
\sum _{\left\{ S_{1},...,S_{n}\right\} _{\text {comp}}}\prod _{i=1}^{n}K_{p}(S_{i})
\end{equation}
 Here \( \left\{ S_{1},...,S_{n}\right\} _{\text {comp}} \) 
denotes
compatible families of polymers, that is \\ 
 \( d(S_{i}^{\ast },S_{j}^{\ast })>1 \) 
for every two \( S_{i} \)
and \( S_{j} \) in the family: recall that by definitions of contours
a polymer \( S \) is a set of columns whose graph that joins all
the points of the dual of the columns of \( S \) at distance \( d(i,j)\leq 1 \)
is connected.

Next, we introduce the so-called truncated contour models defined
with the help of the following
\begin{definition}
A truncated contour functional is defined as
\begin{equation}
\label{eq:3.27}
K_{p}^{\prime }(S)=\left\{ \begin{array}{ll}
K_{p}(S) & \text {if}\, \left\Vert K_{p}(S)\right\Vert 
\leq e^{-\alpha \left\Vert S\right\Vert }\text {}\\
0 & \text {otherwise}
\end{array}\right.
\end{equation}
 where 
\( 
\left\Vert K_{p}(S)\right\Vert 
=\sum _{\Gamma ^{p}:\text {supp}\, \Gamma ^{p}=S}\left| z(\Gamma ^{p})\right|  
\), and \( \alpha >0 \) 
is some positive parameter to be chosen later
(see Theorem~\ref{T:unicity} below).
\end{definition}
\begin{definition}
The collection \( \left\{ S,p\right\}  \) of all \( p \)-contours
\( \Gamma ^{p} \)with support \\
 supp\( \, \Gamma ^{p}=S \) is called stable if
\begin{equation}
\label{eq:3.28}
\left\Vert K_{p}(S)\right\Vert \leq e^{-\alpha \left\Vert S\right\Vert }
\end{equation}
 i.e. if 
\( K_{p}(S)=K_{p}^{\prime }(S) \).
\end{definition}

We define the truncated partition function \( \Xi _{p}^{\prime }(\Omega ) \)
as the partition function obtained from \( \Xi _{p}^{\text {dil}}(\Omega ) \)
by leaving out unstable collections of contours, namely
\begin{eqnarray}
\Xi _{p}^{\prime }(\Omega ) 
& = & 
q^{-e_{p}\left\Vert \Omega \right\Vert }\sideset {}{'}
\sum _{\{\Gamma _{1}^{p},\ldots ,\Gamma _{n}^{p}\}_{\text {comp}}}
\prod _{i=1}^{n}z(\Gamma _{i}^{p})
\\
 & = & q^{-e_{p}\left\Vert \Omega \right\Vert }
\sum _{\left\{ S_{1},...,S_{n}\right\} _{\text {comp}}}
\prod _{i=1}^{n}K_{p}^{\prime }(S_{i})
\label{eq:3.29}
\end{eqnarray}

Here the sum goes over compatible families of 
\textit{stable collections of contours}. 
Let 
\begin{equation}
\label{eq:3.31}
h_{p}=-\lim _{\Omega \rightarrow L}\frac{1}{\left\Vert \Omega \right\Vert }
\ln \Xi _{p}^{\prime }(\Omega )
\end{equation}
 be the \textit{metastable free energy} of the truncated partition
function \( \Xi _{p}^{\prime }(\Omega ) \).

For \( \alpha  \) large enough, the thermodynamic limit (\ref{eq:3.31})
can be controlled by a convergent cluster expansion. 
We conclude the
existence of \( h_{p} \), together with the bounds
\begin{eqnarray}
e^{-\kappa e^{-\alpha }\left| \partial _{s}\Omega \right| } 
& \leq  & \Xi _{p}^{\prime }(\Omega )e^{h_{p}\left\Vert \Omega \right\Vert }
\leq e^{\kappa e^{-\alpha }\left| \partial _{s}\Omega \right| }
\label{eq:3.33} \\
\left| h_{p}-e_{p}\ln q\right|  & \leq  & \kappa e^{-\alpha }
\label{eq:3.34}
\end{eqnarray}
 where 
\( \kappa =\kappa _{\text {cl}}(\chi ^{\prime })^{2} \) 
where
\( \kappa _{\text {cl}}=\frac{\sqrt{5}+3}{2}e^{\frac{2}{\sqrt{5}+1}} \)
is the cluster constant \cite{KP} 
and 
\( \kappa ^{\prime }=3^{d-1}-1 \);
\( \partial _{s}\Omega =\partial \Omega \cap \mathbb {L}_{0} \) in
the way defined in Subsection \ref{S:1.1}.

\begin{theorem}
\label{T:unicity}
Assume that 
\( 1/d<\beta _{b}<1 \) 
and \( q \) 
is large enough so
that 
\( e^{-\alpha }\equiv e^{-\tau +2\kappa ^{\prime }+3}<\frac{0.7}{\kappa \kappa ^{\prime }} \),
then there exists a unique 
\( \beta _{s}^{t}=\frac{1}{d-1}(1-\beta _{b})+O(e^{-\tau }) \)
such that :
\begin{description}
\item [(i)]
for 
\( \beta _{s}=\beta _{s}^{t} \)
\[
\Xi _{p}^{\text {dil}}(\Omega )=\Xi _{p}^{\prime }(\Omega )
\]
 for both boundary conditions \( p= \)o and \( p= \)of, and the
free energy of the hydra model is given by 
\( g_{\text {f}}+\left[ (d-1)\beta _{s}+\beta _{b}\right] \ln q=h_{\text {o}}=h_{\text {of}} \)
\item [(ii)]
for \( \beta _{s}>\beta _{s}^{t} \) 
\[
\Xi _{\text {o}}^{\text {dil}}(\Omega )=\Xi _{\text {o}}^{\prime }(\Omega )
\]
 and 
\( g_{\text {o}}+\left[ (d-1)\beta _{s}+\beta _{b}\right] \ln q=h_{\text {o}}<h_{\text {of}} \)
\item [(iii)]
for \( \beta _{s}<\beta _{s}^{t} \)
\[
\Xi _{\text {of}}^{\text {dil}}(\Omega )=\Xi _{\text {of}}^{\prime }(\Omega )
\]
 and 
\( g_{\text {of}}+\left[ (d-1)\beta _{s}+\beta _{b}\right] \ln q=h_{\text {fo}}<h_{\text {f}} \)
\end{description}
\end{theorem}

\begin{proof}
Starting from the Peierls estimates given in Proposition~\ref{P:Peierls}
and the definitions of this subsection the proof is the same as that
of Theorem~3.5 in paper I. 
Let us only recall that to exponentiate
the partition function 
\( \mathcal{Z}_{p}(\Omega )=q^{e_{p}\left\Vert \Omega \right\Vert }\Xi _{p}^{\prime }(\Omega ) \)
we define the truncated functional 
\( \Phi ^{T} \) associated to
\( K_{p}^{\prime } \)
\begin{equation}
\Phi ^{T}(X)=\frac{a(X)}{\prod _{\gamma }X(S)!}\prod _{S}K_{p}^{\prime }(S)^{X(S)}
\end{equation}
 defined on the multi-indexes \( X \) associated to the polymers
(a multi-index being a function from the set of polymers into the
set of non negative integers, and \( a(X) \) is defined as in (\ref{eq:2.11})).
The number of polymers \( S \) with number of columns 
\( \left\Vert S\right\Vert =n \)
and containing a given column can be bounded by \( \nu ^{n} \) where
\( \nu =(3^{d-1}-1)^{2} \) as in paper I: this is because the chosen
definition for connectedness of columns here is the same as that for
connectedness of lines in paper I.

As a result of the standard cluster expansion \cite{KP,M}, we get
for \( \kappa e^{-\alpha }<1 \) 
\[
\mathcal{Z}_{p}(\Omega )=\exp \left\{ \sum _{X}\Phi^{T}(X)\right\} 
\]
where the sum is over multi-indexes whose support 
\( \text {supp}X=\left\{ S:X(S) \geq 1\right\}  \)
belongs to \( \Omega  \). 
The series 
\( \sum _{X:\text {supp}X\backepsilon L}\left| \Phi^{T}(X)\right|  \)
is absolutely convergent and satisfies the bound 
\begin{equation}
\sum _{X:\text {supp}X\backepsilon L}\left| \Phi ^{T}(X)\right| \leq \kappa e^{-\alpha }
\end{equation}
\end{proof}

Let us introduce the Gibbs states \( \langle \, \cdot \, \rangle ^{p} \),
associated to the boundary conditions \( p\in \{ \)f\( ; \)of\( \} \).
Theorem above show also that at \( \beta _{s}=\beta _{s}^{t} \)
\begin{eqnarray}
\langle \, n_{b}\, \rangle ^{\text {o}} & = & 1-O(e^{-\tau })
\label{eq:3.41} 
\\
\langle \, n_{b}\, \rangle ^{\text {of}} & = & O(e^{-\tau })
\label{eq:3.42}
\end{eqnarray}
for any bond \( b \) of the boundary layer and any bond \( b \)
between the boundary layer and the first layer.

Indeed by the correspondence (\ref{eq:DA2}) these equations are equivalent
to
\begin{eqnarray}
\langle \, \widehat{n}_{b^{\ast }}\, \rangle ^{\text {o}} & = & O(e^{-\tau })
\label{eq:P1} 
\\
\langle \, \widehat{n}_{b^{\ast }}\, \rangle ^{\text {of}} & = & 1-O(e^{-\tau })
\label{eq:P2}
\end{eqnarray}
for the dual cells of the bonds under consideration. 
By definition
of contours, with ordered (\( \text {o} \)) boundary conditions,
any such cells are occupied only if there is an ordered contour surrounding
it and that the correlation functions are controlled by the contour
model cluster expansion. With ordered-free (\( \text {of} \)) boundary
conditions, such cells are empty only if there is a \( \text {of} \)--contour
surrounding it and again the correlations are controlled by cluster
expansion. Obviously, the relation (\ref{eq:P1}) holds true for any
\( \beta _{s}\geq \beta _{s}^{t} \) while the relation (\ref{eq:P2})
hold true for any \( \beta _{s}\leq \beta _{s}^{t} \).

This shows in particular that the derivative 
\( \frac{\partial }{\partial K}g_{\text {o}} \)
of the free energy \( g_{\text {o}} \) 
with respect to the surface
coupling constant \( K \) is discontinuous near 
\( K=\beta ^{-1}\ln \left( 1+\left( \frac{q}{e^{\beta J}-1}\right) ^{1/(d-1)}\right)  \).

\subsection*{Acknowledgments}

The authors thank S.\ Miracle-Sol\'{e}, S.\ Shlosman, and V.\ Zagrebnov
for helpful discussions. L.L.\ acknowledges the warm hospitality and
financial support of the Centre de Physique Th\'{e}orique.
This work (L.\ L.) have been partially supported by CNRST-Morocco (PARS.\ 037).

\noindent
\textit{Note added after publication:}
Let us mention Refs.\ \cite{CISS}
in which, besides numerical results analytical calculations 
in the large-q limit, mainly in 2d, are presented.

\section*{Appendix: Proof of Proposition \ref{P:Peierls}}

\renewcommand{\theequation}{A.\arabic{equation}} 
\renewcommand{\thesection}{A} 
\setcounter{equation}{0} 
\setcounter{theorem}{0}

We begin the proof by considering contours 
\( \Gamma =\left\{ \text {supp}\, \Gamma ,\widehat{X}^{\Gamma },\widehat{Y}^{\Gamma }\right\}  \)
(where 
\( \left\{ \widehat{X}^{\Gamma },\widehat{Y}^{\Gamma }\right\}  \)
is the configuration having \( \Gamma  \) 
as unique contour) without
polymers, i.e.\ 
\( \widehat{Y}^{\Gamma }=\emptyset  \). 
We have the
decomposition 
\( \widehat{X}^{\Gamma }=\widehat{X}_{s}^{\Gamma }\cup  \)
\( \widehat{Z}_{b}^{\Gamma }\cup \widehat{Y}_{b}^{\Gamma } \) 
where
\( \widehat{X}_{s}^{\Gamma }
=\widehat{X}^{\Gamma }\cap \left[ B(\mathbb {L}_{0})\right] ^{\ast } \),
\( \widehat{Y}_{b}^{\Gamma }
=\widehat{X}^{\Gamma }\cap \left[ B(\mathbb {L\setminus L}_{0})\right] ^{\ast } \),
and 
\( 
\widehat{Z}_{b}^{\Gamma }
=\widehat{X}^{\Gamma }\setminus (\widehat{X}_{s}^{\Gamma }\cup \widehat{Y}_{b}^{\Gamma }) 
\).

A \( d \)--cell \( x^{\ast }\in \mathbb {L}_{0}^{\ast } \) 
will
be called regular if the \( \left( d-1\right)  \)--cells of its boundary
that belong to the boundary layer are either all empty or all occupied.
It will be called irregular otherwise. 
We denote by \( R_{0}(\Gamma ) \)
the set of correct \( d \)--cells 
of the contour \( \Gamma  \):
\( R_{0}(\Gamma )=\left\{ x^{\ast }\in \mathbb {L}_{0}:\left| X^{\Gamma }
\cap \mathcal{E}(x^{\ast })\right| \, \text {equals}\, 0\, \text {or}\, 2d-1\right\}  \).
We let \( I_{0}(\Gamma ) \) 
be the set of incorrect \( d \)--cells
of the contour 
\( \Gamma  \)
: 
\( 
I_{0}(\Gamma )
=\left\{ 
x^{\ast }\in \mathbb {L}_{0}:1\leq \left| X^{\Gamma }\cap \mathcal{E}(x^{\ast })\right| 
\leq 2(d-1)\right\}  
\).

\begin{lemma}
\label{L:AP1}

\begin{equation}
\label{eq:A.1}
\varrho (\Gamma )q^{\underline{e}\left\Vert \Gamma \right\Vert }
\leq q^{-\frac{1-\beta _{b}}{2(d-1)}\left| I_{0}(\Gamma )\right| 
-\left( \beta _{b}-\frac{1}{d}\right) \left| Y_{b}^{\Gamma }\right| }
\end{equation}
\end{lemma}

\begin{proof}
By Lemma \ref{L:I1} and definition (\ref{eq:3.9}), 
one has
\[
\varrho (\Gamma )
=q^{\sum _{x^{\ast }\in R_{0}(\Gamma )\cup I_{0}(\Gamma )}
e_{x^{\ast }}(\widehat{X})-\beta _{b}\left| \widehat{Y}_{b}\right| 
+\widetilde{N}_{\text {cl}}(\widehat{X})}
\]
where 
\( 
e_{x^{\ast }}(\widehat{X})
=-\frac{\beta _{s}}{2}\left| \widehat{X}_{s}\cap \mathcal{E}(x^{\ast })\right| 
-\beta _{b}\left| \widehat{Z}_{b}\cap \mathcal{E}(x^{\ast })\right|  
\)
and to simplify formulae we put hereafter 
\( \widehat{X} \), \( \widehat{X}_{s} \), \( \widehat{Z}_{b} \)
and \( \widehat{Y}_{b} \) 
instead of \( \widehat{X}^{\Gamma } \),
\( \widehat{X}_{s}^{\Gamma } \), \( \widehat{Z}_{b}^{\Gamma } \)
and \( \widehat{Y}_{b}^{\Gamma } \). 
We define 
\[
\overline{N}_{\text {cl}}(\widehat{X}_{s}\cup \widehat{Z}_{b})
=\widetilde{N}_{\text {cl}}(\widehat{X}_{s}\cup \widehat{Z}_{b})-
\sum _{x^{\ast }\in R_{0}(\Gamma )}\widetilde{N}_{\text {cl}}(\widehat{X}\cap \mathcal{E}(x^{\ast }))
\]
 as the number of independent closed surfaces that are not boundaries
of an occupied \( d \)--cell of the surface. 
This leads to the decomposition
\begin{equation}
\label{eq:A.5}
\varrho (\Gamma )q^{\underline{e}\left\Vert \Gamma \right\Vert }
=q^{-\mathbf{A}_{s}(\Gamma )-\mathbf{B}_{s}(\Gamma )-\mathbf{A}_{b}(\Gamma ^{p})}
\end{equation}
 where
\begin{eqnarray}
\mathbf{A}_{s}(\Gamma ) 
& = & \sum _{x^{\ast }\in R_{0}(\Gamma )}
\left[ e_{x^{\ast }}(\widehat{X})
-\underline{e}-\widetilde{N}_{\text {cl}}(\widehat{X}\cap \mathcal{E}(x^{\ast }))\right] 
\label{eq:A.6} \\
\mathbf{B}_{s}(\Gamma ) & = & \sum _{x^{\ast }\in I_{0}(\Gamma )}\left[ e_{x^{\ast }}(\widehat{X})
-\underline{e}\right] -\overline{N}_{\text {cl}}(\widehat{X}_{s}\cup \widehat{Z}_{b})
\label{eq:A.3} \\
\mathbf{A}_{b}(\Gamma ) 
& = & \beta _{b}\left| \widehat{Y}_{b}\right| 
-\left[ \widetilde{N}_{\text {cl}}(\widehat{X})-\widetilde{N}_{\text {cl}}(\widehat{X}_{s}
\cup \widehat{Z}_{b})\right] 
\label{eq:A.7}
\end{eqnarray}
 Clearly 
\begin{equation}
\label{eq:A.10}
\mathbf{A}_{s}(\Gamma )\geq 0
\end{equation}
 Indeed the regular \( d \)--cells are either empty in which case
the term inside brackets in (\ref{eq:A.6}) equals 
\[
-\underline{e}
=-\min \left\{ 
e_{\text {o}}=0;e_{\text {of}}=(d-1)\beta _{s}+\beta _{b}-1
\right\} 
\]
 or they are occupied in which case this term equals 
\[
(d-1)\beta _{s}+\beta _{b}-1
-\min \left\{ e_{\text {o}}=0; e_{\text {of}}=(d-1)\beta _{s}+\beta _{b}-1\right\} 
\]

Let us now bound \( \mathbf{B}_{s}(\Gamma ) \). 
We first notice that
for incorrect \( d \)--cells \( x^{\ast } \) of the surface,
\begin{eqnarray}
e_{x^{\ast }}(\widehat{X})-\underline{e} 
& = & \frac{\beta _{s}}{2}\left| \widehat{X}_{s}\cap \mathcal{E}(x^{\ast })\right| 
+\beta _{b}\left| \widehat{Z}_{b}\cap \mathcal{E}(x^{\ast })\right| 
-e_{\text {of}}\chi \left( \beta _{s}\leq \frac{1-\beta _{b}}{d-1}\right) 
\nonumber \\
 & \geq  & 
\frac{1-\beta _{b}}{2(d-1)}\left| \widehat{X}_{s}\cap \mathcal{E}(x^{\ast })\right| 
+\beta _{b}\left| \widehat{Z}_{b}\cap \mathcal{E}(x^{\ast })\right| 
\label{eq:A.8}
\end{eqnarray}
Furthermore the number 
\( \overline{N}_{\text {cl}}(\widehat{X}_{s}\cup \widehat{Z}_{b}) \)
may be bounded as
\begin{equation}
\label{eq:A.9}
\overline{N}_{\text {cl}}(\widehat{X}_{s}\cup \widehat{Z}_{b})
\leq 
\sum _{x^{\ast }:d-1\leq \left| \widehat{X}\cap \mathcal{E}(x^{\ast })\right| 
\leq 2(d-1)-1\atop \left| \widehat{Z}_{b}\cap \mathcal{E}(x^{\ast })\right| =1}
\frac{2^{\left| \widehat{X}\cap \mathcal{E}(x^{\ast })\right| }}{2^{2(d-1)}}
\leq 
\sum_{x^{\ast }:d-1 \leq \left| \widehat{X}\cap \mathcal{E}(x^{\ast })\right| 
\leq 2(d-1)-1\atop \left| \widehat{Z}_{b}\cap \mathcal{E}(x^{\ast })\right| =1}
\frac{1}{2}
\end{equation}

If an incorrect site of the surface is such that 
\( \left| \widehat{Z}_{b}\cap \mathcal{E}(x^{\ast })\right| =0 \),
then necessarily \( \left| \widehat{X}_{s}\cap \mathcal{E}(x^{\ast })\right|=1  \)
and thus such site gives a contribution 
\[
\frac{1-\beta _{b}}{2(d-1)}
\]
 to \( \mathbf{B}_{s}(\Gamma ) \). 
Let us now consider those incorrect
\( d \)--cells for which \( \left| \widehat{Z}_{b}\cap \mathcal{E}(x^{\ast })\right| =1 \).
Starting from (\ref{eq:A.8}) we get for such cells
\begin{eqnarray*}
e_{x^{\ast }}(\widehat{X})-\underline{e} 
& \geq  & \beta _{b}+\frac{1-\beta _{b}}{2(d-1)}\left| \widehat{X}_{s}\cap \mathcal{E}(x^{\ast })\right| 
\\
 & = & 
\frac{1-\beta _{b}}{2(d-1)}+\beta _{b}\frac{2(d-1)+1-\left| \widehat{X}_{s}
\cap \mathcal{E}(x^{\ast })\right| }{2(d-1)}
+\frac{\left| \widehat{X}_{s}\cap \mathcal{E}(x^{\ast })\right| -1}{2(d-1)}
\\
 & \geq  & \frac{1-\beta _{b}}{2(d-1)}
+\frac{1}{d}\frac{2(d-1)+1-\left| \widehat{X}_{s}\cap \mathcal{E}(x^{\ast })\right| }{2(d-1)}
+\frac{\left| \widehat{X}_{s}\cap \mathcal{E}(x^{\ast })\right| -1}{2(d-1)}
\\
 & = & 
\frac{1-\beta _{b}}{2(d-1)}
+\frac{\left| \widehat{X}_{s}\cap \mathcal{E}(x^{\ast })\right| +1}{2d}
\end{eqnarray*}
 where for the second inequality we take into account that 
\( \left| \widehat{X}_{s}\cap \mathcal{E}(x^{\ast })\right| \leq2 (d-1) \)
and 
\( \beta _{b}\geq1 /d \). 
When furthermore 
\( \left| \widehat{X}_{s}\cap \mathcal{E}(x^{\ast })\right| \geq d-1 \),
one infer 
\[
e_{x^{\ast }}(\widehat{X})-\underline{e}\geq \frac{1-\beta _{b}}{2(d-1)}+\frac{1}{2}
\]
 and it thus follows from (\ref{eq:A.3}), (\ref{eq:A.9}),  
and (\ref{eq:A.8}),
that each incorrect cell with 
\( \left| \widehat{Z}_{b}\cap \mathcal{E}(x^{\ast })\right| =1 \),
gives also a contribution at least \( \frac{1-\beta _{b}}{2(d-1)} \)
to 
\( \mathbf{B}_{s}(\Gamma ) \). 
Therefore 
\begin{equation}
\label{eq:A.11}
q^{-\mathbf{B}_{s}(\Gamma )}
\leq q^{-\frac{1-\beta _{b}}{2(d-1)}\left| I_{0}(\Gamma )\right| }
\end{equation}
 Consider finally, the quantity \( \mathbf{A}_{b}(\Gamma ) \). 
We
will prove the inequality 
\begin{equation}
\label{eq:A.12}
\widetilde{N}_{\text {cl}}
(\widehat{X})-\widetilde{N}_{\text {cl}}(\widehat{X}_{s}\cup \widehat{Z}_{b})
\leq 
\frac{\left| \widehat{Y}_{b}\right| }{d}
\end{equation}
 Notice first the obvious inequality
\[
\widetilde{N}_{\text {cl}}
(\widehat{X})-\widetilde{N}_{\text {cl}}(\widehat{X}_{s}\cup \widehat{Z}_{b})
\leq N_{\text {cl}}(\widehat{Y}_{b}\cup B_{01}^{\ast })
\]
 where \( B_{01} \) is the set of bonds between the boundary layer
and the first layer and 
\( N_{\text {cl}}(\widehat{Y}_{b}\cup B_{01}^{\ast }) \)
is the number of independent closed surfaces of 
\( \widehat{Y}_{b}\cup B_{01}^{\ast } \).
The number \( \left| \widehat{Y}_{b}\right|  \) 
can be written
\[
\left| \widehat{Y}_{b}\right| 
=
\sum _{x^{\ast }\in \left[ \mathbb {L}\diagdown \mathbb {L}_{0}\right] ^{\ast }
:\left| \widehat{Y}_{b}\cap \mathcal{E}(x^{\ast })\right| \geq 1}
\frac{\left| \widehat{Y}_{b}\cap \mathcal{E}(x^{\ast })\right| }{2}
\]
 (because each \( (d-1) \)--cell belongs to the boundary of two \( d \)--cells).

For the configurations \( \widehat{Y}_{b} \) that do not hit \( B_{01}^{\ast } \)
(meaning that there are no cell of \( \widehat{Y}_{b} \) connected
with \( B_{01}^{\ast } \) in the \( \mathbb {R}^{d} \) sense) we
get immediately (\ref{eq:A.12}) as already used in the proof of Theorem~\ref{T:CE}:
closed surfaces of minimal area are \( d \)--cells and the number
of \( (d-1) \)--cells in the boundary of \( d \)--cell equals \( 2d \).

For the configurations \( \widehat{Y}_{b} \) that do hit \( B_{01}^{\ast } \),
notice first that \( \left| \widehat{Y}_{b}\right|  \) can be written
\begin{equation}
\label{eq:AA}
\left| \widehat{Y}_{b}\right| 
=\sum _{\substack {x^{\ast }\in \left[ \mathbb {L}\diagdown \mathbb {L}_{0}\right] ^{\ast }
:\left| \widehat{Y}_{b}\cap \mathcal{E}(x^{\ast })\right| \geq 2}}
\frac{\left| \widehat{Y}_{b}\cap \mathcal{E}(x^{\ast })\right| }{2}
+\sum _{\substack {x^{\ast }\in \left[ \mathbb {L}\diagdown \mathbb {L}_{0}\right] ^{\ast }
:\left| \widehat{Y}_{b}\cap \mathcal{E}(x^{\ast })\right| =1}}
\frac{\left| \widehat{Y}_{b}\cap \mathcal{E}(x^{\ast })\right| }{2}
\end{equation}
 Since \( \widehat{Y}_{b} \) 
is finite, the set of cells with 
\( \left| \widehat{Y}_{b}\cap \mathcal{E}(x^{\ast })\right| =1 \)
is non empty and furthermore for any 
\( (d-1) \)--cell of \( B_{01}^{\ast } \)
there exists a \( d \)--cell \( x^{\ast } \)
above it such that 
\( \left| \widehat{Y}_{b}\cap \mathcal{E}(x^{\ast })\right| =1 \)
that can not contribute to 
\( N_{\text {cl}}(\widehat{Y}_{b}\cup B_{01}^{\ast }) \).
Now closed surfaces of minimal area are \( d \)--cells having in
their boundary \( 2d-1 \) cells of \( \widehat{Y}_{b} \) and a cell
of \( B_{01}^{\ast }\). 
Thus for such surfaces we have a contribution
\( d-1/2 \) coming from the first term of the RHS of (\ref{eq:AA})
and a contribution \( 1/2 \) coming from the second term.

This implies (\ref{eq:A.12}) giving
\begin{equation}
\label{eq:A.14}
q^{-\mathbf{A}_{b}(\Gamma )}
\leq q^{-\left( \beta _{b}-\frac{1}{d}\right) \left| \widehat{Y}_{b}\right| }
\end{equation}
 which in turn implies the lemma by taking into account (\ref{eq:A.5}),
(\ref{eq:A.10}) and (\ref{eq:A.11}).
\end{proof}

Considering still contours 
\( \Gamma =\left\{ \text {supp}\Gamma ,X^{\Gamma },Y^{\Gamma }\right\}  \)
without polymers, i.e.\ (\( Y^{\Gamma }=\emptyset  \)) 
we have the

\begin{lemma}\label{L:AP2}
Assume that 
\( \beta _{b}>\frac{1}{d} \), and \( 2\nu _{d}q^{\frac{1}{d}-\beta _{b}}<1 \),
then 
\begin{equation}
\label{eq:A.15}
\sum _{\Gamma :\text {supp}\, \Gamma =S}\varrho (\Gamma )
q^{\underline{e}\left\Vert \Gamma \right\Vert }
\leq \left( 2^{(3d-2)}q^{^{\frac{1-\beta _{b}}{2(d-1)}}-}
+2^{d+1}\nu _{d}q^{\frac{1}{d}-\beta _{b}}\right) ^{\left\Vert S\right\Vert }
\frac{1}{1-2\nu _{d}q^{\frac{1}{d}-\beta _{b}}}
\end{equation}
\end{lemma} 
which shows that, whenever \( q \) is large enough, the
Peierls condition holds true for the class of contours without polymers.

\begin{proof}

First, observe that for contours \( \Gamma  \) with support 
\(\text{supp} \, \Gamma =S \)
and number of irregular cells of the boundary layer 
\( \left| I_{0}(\Gamma )\right| =k \)
one has 
\( 
\left| \widehat{Y}_{b}\right| 
=\left| \delta _{1}\right| +...+\left| \delta _{m}\right| \geq \left\Vert S\right\Vert -k 
\).
Therefore,
\begin{eqnarray}
\sum _{\Gamma :\text {supp}\, \Gamma =S}\varrho (\Gamma )
q^{\underline{e}\left\Vert \Gamma \right\Vert } 
& \leq  & 
\sum _{0\leq k\leq \left\Vert S\right\Vert }
\sum _{\Gamma ^{p}:\left| I_{0}(\Gamma )\right| =k}
q^{-\frac{1-\beta _{b}}{2(d-1)}k}
q^{(\frac{1}{d}-\beta _{b})\left| \widehat{Y}_{b}\right| }
\nonumber 
\\
 & \leq  & 
\sum _{0\leq k\leq \left\Vert S\right\Vert }
\binom{\left\Vert S\right\Vert }{k}2^{(2d-1)k}2^{\left\Vert S\right\Vert -k}
q^{-\frac{1-\beta _{b}}{2(d-1)}k}
\\
 &  & \times 
\sum _{n\leq 2\left\Vert S\right\Vert }
\sum _{\delta _{1}\ni  s_{1},...,\delta _{n} \ni  s_{n}
\atop \left| \delta _{1}\right| +...+\left| \delta _{n}\right| \geq \left\Vert S\right\Vert -k}
\sum _{s_{1},...,s_{n}\atop s_{\alpha }\in S;s_{\alpha }\nsim B_{01}^{\ast }}
\prod _{j=1}^{m}q^{\left( \frac{1}{d}-\beta _{b}\right) \left| \delta _{j}\right| }
\nonumber
\end{eqnarray}
Here the binomial coefficient 
\( \binom{\left\Vert S\right\Vert }{k} \)
bounds the choice of irregular cells of the dual of the boundary layer
while the factor 
\( 2^{(2d-1)k}2^{\left\Vert S\right\Vert -k} \)
bounds the numbers of contours with \( \left\Vert S\right\Vert  \)
columns and \( k \) irregular cells; 
the notation \( s_{\alpha }\nsim B_{01}^{\ast } \)
means that a \( (d-2) \)--cell of the boundary of the \( (d-1) \)--cell
\( s_{\alpha } \) belongs to the boundary \( \mathcal{E}(B_{01}^{\ast }) \).
Then
\begin{eqnarray}
 \sum _{\Gamma :\text {supp}\, \Gamma =S}\varrho (\Gamma )
q^{\underline{e}\left\Vert \Gamma \right\Vert }
&\leq &  
\sum _{0\leq k\leq \left\Vert S\right\Vert }
\binom{\left\Vert S\right\Vert }{k}
2^{(2d-1)k}
2^{\left\Vert S\right\Vert -k}
q^{-\frac{1-\beta _{b}}{2(d-1)}k}
\\
 &   \times &
\sum _{n\leq 2\left\Vert S\right\Vert }\binom{(d-1)\left\Vert S\right\Vert }{n}
\sum _{m_{1}+...+m_{n}
\geq 
\left\Vert S\right\Vert -k}\prod _{j=1}^{n}\left( \nu _{d}q^{\frac{1}{d}-\beta _{b}}\right) ^{m_{j}}
\nonumber 
\end{eqnarray}
Here the binomial coefficient 
\( \binom{(d-1)\left\Vert S\right\Vert }{n} \)
bounds the choice for the components \( \delta _{1},...,\delta _{n} \)
of \( Y_{b} \) to hit the boundary layer at \( s_{1},...,s_{n} \).
The above inequality yields
\begin{eqnarray}
 \sum _{\Gamma :\text {supp}\, \Gamma =S}\varrho (\Gamma )
q^{\underline{e}\left\Vert \Gamma \right\Vert }
&\leq & \sum _{0\leq k\leq \left\Vert S\right\Vert }
\binom{\left\Vert S\right\Vert }{k}2^{(2d-1)k}2^{\left\Vert S\right\Vert -k}
q^{-\frac{1-\beta _{b}}{2(d-1)}k}
\nonumber \\
 &  & \hphantom {xx}\times 
\sum _{n\leq 2\left\Vert S\right\Vert }\binom{(d-1)\left\Vert S\right\Vert }{n}
\sum _{m\geq \left\Vert S\right\Vert -k}\left( 2\nu _{d}q^{\frac{1}{d}-\beta _{b}}\right) ^{m}
\nonumber 
\\
&\leq& 
\sum _{0\leq k\leq \left\Vert S\right\Vert }
\binom{\left\Vert S\right\Vert }{k}2^{(2d-1)k}2^{\left\Vert S\right\Vert -k}
q^{-\frac{1-\beta _{b}}{2(d-1)}k}
\nonumber 
\\
 &  & \hphantom {xx}
\times \left( 2\nu _{d}
q^{-\left( \frac{1}{d}-\beta _{b}\right) }\right) ^{\left\Vert S\right\Vert -k}
\frac{2^{(d-1)\left\Vert S\right\Vert }}{1-2\nu _{d}q^{\frac{1}{d}-\beta _{b}}}
\end{eqnarray}
 that gives the inequality of the lemma.
\end{proof}

We now turn to the general case of contours with non empty polymers
and first give a bound on the activity \( \psi \left( \gamma \right)  \)
of polymers.

\begin{lemma}
\label{L:AP3}
Assume that \( \beta _{b}>\frac{1}{d} \), 
and 
\( c\nu _{d}^{2}q^{-\frac{1}{d}-\beta _{b}}\leq 1 \)
with 
\( c=8e(e-1)c_{0} \) and \( \nu _{d}=(2d)^{2} \), 
then
\begin{equation}
\left| \psi \left( \gamma \right) \right| 
\leq \left( c\nu _{d}^{2}q^{\frac{1}{d}-\beta _{b}}\right) ^{\left| \gamma \right| }
\end{equation}
\end{lemma}

\begin{proof}
Let us first recall the definition (\ref{eq:3.6}): 
\( \psi (\gamma )\equiv \sum _{A:\text {supp}\, A=\gamma }\omega (A) \)
where the weights of aggregates are defined by (see (\ref{eq:3.3})
and (\ref{eq:3.4})): 
\( \omega (A)=\prod _{\gamma \in A}(e^{-\Phi (\gamma )}-1) \).
By Theorem~\ref{T:CE} we know that 
\( \left| \Phi (\gamma )\right| 
\leq \left( ec_{0}\nu _{d}q^{\frac{1}{d}-\beta _{b}}\right) ^{\left| \gamma \right| }(\leq 1) \)
for \( q \) 
large enough. 
Since for any 
\( \left| x\right| \leq 1 \),
\( \left| e^{-x}-1\right| \leq (e-1)\left| x\right|  \), 
we have
\begin{equation}
\left| \Psi (\gamma )\right| 
=\left| e^{-\Phi (\gamma )}-1\right| 
\leq (e-1)\left| \Phi (\gamma )\right| 
\leq 
\left( (e-1)ec_{0}\nu _{d}q^{\frac{1}{d}-\beta _{b}}\right) ^{\left| \gamma \right| }
\equiv e^{-\sigma \left| \gamma \right| }
\end{equation}

Then,
\begin{eqnarray}
\sum _{A:\text {supp}\, A=\gamma }\left| \omega (A)\right|  
& = & 
\sum _{n\geq 1}
\sum _{\gamma _{1},...,\gamma _{n}: \atop 
\text {supp}\, \left\{ \gamma _{1},...,\gamma _{n}\right\} =\gamma }
\prod _{j=1}^{n}\left| \Psi (\gamma _{j})\right| 
\nonumber 
\\
 &  & \leq \sum _{n\geq 1}2^{\left| \gamma \right| }
\sum _{
\gamma _{1}\ni s_{1},...,
\gamma _{n}\ni s_{n}: \atop \text {supp}\, 
\left\{ \gamma _{1},...,\gamma _{n}\right\} =\gamma 
}
\prod _{j=1}^{n}e^{-\sigma \left| \gamma _{j}\right| }
\nonumber 
\\
 &  & 
\leq 
\sum _{n\geq 1}2^{\left| \gamma \right| }
\sum _{\substack {m_{1},...,m_{n}:\atop m_{1}+...+m_{n}\geq \left| \gamma \right| }}
\prod _{j=1}^{n}\left( \nu _{d}e^{-\sigma }\right) ^{m_{j}}
\nonumber \\
 &  & 
\leq 
\sum _{n\geq 1}\sum _{\substack {m_{1},...,m_{n}:\atop m_{1}+...+m_{n}\geq \left| \gamma \right| }}
\prod _{j=1}^{n}\left( 2\nu _{d}e^{-\sigma }\right) ^{m_{j}}
\end{eqnarray}
Here, we used as in the proof of Theorem~\ref{T:CE}
that the number of polymers of length \( m \) containing a given
bond or a given vertex is less than 
\( \nu _{d}^{m} \); 
the term
\( 2^{\left| \gamma \right| } \) bounds the combinatoric choice of
the cells \( s_{j}\in \gamma _{j} \), because \( \gamma  \) being
connected, it contains \( n-1 \) such intersecting cells (see \cite{GMM}).

We put 
\( k=m_{1}+...+m_{n} \) 
and notice that there are at most
\( \binom{k}{n-1} \) 
such numbers to get
\begin{eqnarray}
\sum _{A:\text {supp}\, A=\gamma }
\left| \omega (A)\right|  
& = & 
\sum _{1\leq n\leq k}\sum _{k\geq \left| \gamma \right| }
\binom{k}{n-1}\left( 2\nu _{d}e^{-\sigma }\right) ^{k}
\nonumber 
\\
 & \leq  & 
\sum _{k\geq \left| \gamma \right| }\left( 4\nu _{d}e^{-\sigma }\right) ^{k}
=\sum _{k\geq \left| \gamma \right| }
\left( \frac{c}{2}\nu _{d}^{2}q^{\frac{1}{d}-\beta _{b}}\right) ^{k}
\nonumber 
\\
 & \leq  & 
\frac{1}{1-\frac{c}{2}\nu _{d}^{2}q^{-\frac{1}{d}+\beta _{b}}}
\left( \frac{c}{2}\nu _{d}^{2}
q^{\frac{1}{d}-\beta _{b}}\right) ^{\left| \gamma \right| }
\end{eqnarray}
provided that 
\( \frac{c}{2}\nu _{d}^{2}q^{\frac{1}{d}-\beta _{b}}<1 \).
The lemma then follows by assuming that 
\( \frac{c}{2}\nu _{d}^{2}q^{\frac{1}{d}-\beta _{b}}\leq \frac{1}{2} \).
\end{proof}

We finally turn to the

\textbf{Proof of Proposition \ref{P:Peierls}}

Consider a contour 
\( \Gamma =\left\{ \text {supp}\, \Gamma ,\widehat{X}^{\Gamma },\widehat{Y}^{\Gamma }\right\}  \)
and as above the decomposition 
\( X^{\Gamma }=\widehat{X}_{s}^{\Gamma }\cup \widehat{Z}_{b}^{\Gamma }\cup \widehat{Y}_{b}^{\Gamma } 
\). 
Consider also the union \( \widehat{T}_{b}=\widehat{Y}_{b}^{\Gamma }\cup \widehat{Y}^{\Gamma } \).
Notice that the set 
\( \widehat{T}=\widehat{X}_{s}^{\Gamma }\cup \widehat{Z}_{b}^{\Gamma }\cup \widehat{T}_{b} \)
is a family of hydras and there are at most 
\( 3^{\left| \widehat{T}_{b}\right| } \)
contours corresponding to this family: this is because a 
\( (d-1) \)--cell
in \( \widehat{T}_{b} \) 
may be occupied either by \( \widehat{Y}_{b}^{\Gamma } \)
or by \( \widehat{Y}^{\Gamma } \) 
or by both. 
Let
\begin{equation}
\left| \widetilde{\varrho }(\widehat{T})\right|
=
\sum _{\Gamma :\widehat{Y}_{b}^{\Gamma }\cup \widehat{Y}^{\Gamma }
=
\widehat{T}}\left| \varrho (\widehat{T})\right|
\end{equation}
 The above remark on the number of contours associated to \( \widehat{T} \)
and Lemma~\ref{L:AP3} imply
\begin{eqnarray}
\left| \widetilde{\varrho }(\widehat{T}) \right|
q^{\underline{e}\left\Vert  \Gamma \right\Vert }  
& \leq  & q^{-\frac{\left| I_{0}(\Gamma )\right| }{2(d-1)}}
\left( 3\sup \left\{ q^{\frac{1}{d}-\beta _{b}},c\nu _{d}^{2}
q^{\frac{1}{d}-\beta _{b}}\right\} \right) ^{\left| \widehat{T}_{b}\right| }
\nonumber 
\\
 & \leq  & 
q^{-\frac{\left| I_{0}(\Gamma )\right| }{2(d-1)}}\left( 3c\nu _{d}^{2}
q^{\frac{1}{d}-\beta _{b}}\right) ^{\left| \widehat{T}_{b}\right| }
\end{eqnarray}
 The rest of the proof is then analog to that of Lemma~\ref{L:AP2}
starting from Lemma~\ref{L:AP3} and replacing \( q^{\frac{1}{d}-\beta _{b}} \)
by 
\( 3c\nu _{d}^{2}q^{\frac{1}{d}-\beta _{b}} \). 
It gives
\begin{equation}
\sum _{\Gamma :\text {supp}\, \Gamma =S}
\left| \varrho (\Gamma )\right| 
q^{\underline{e}\left\Vert \Gamma \right\Vert }
\leq \left( 2^{(3d-2)}
q^{-\frac{1-\beta _{b}}{2(d-1)}}+3c2^{d+1}\nu _{d}^{3}
q^{\frac{1}{d}-\beta _{b}}\right) ^{\left\Vert S\right\Vert }
\frac{1}{1-6c\nu _{d}^{3}q^{\frac{1}{d}-\beta _{b}}}
\end{equation}
 provided 
\( 6c\nu _{d}^{3}q^{\frac{1}{d}-\beta _{b}}<1 \) 
and ends
the proof of the proposition. \rule{0.5em}{0.5em}

\bibliographystyle{unsrt}

\end{document}